{{\PassOptionsToPackage
{spanish,english} {babel}
}}

\documentclass{aa}  

\usepackage{babel}
\usepackage[nottoc]{tocbibind} 
\usepackage{graphicx}
\usepackage{csvsimple}
\usepackage{txfonts}
\usepackage{lscape}
\usepackage{natbib}
\usepackage[utf8]{inputenc}
\usepackage[T1]{fontenc}
\bibpunct{(}{)}{;}{a}{}{,}
\usepackage[colorlinks, citecolor=blue]{hyperref}
\usepackage[toc,page]{appendix}
\usepackage{booktabs,caption}
\usepackage[flushleft]{threeparttable}

\newcommand{\hii}{\relax \ifmmode {\mbox H\,{\scshape ii}}\else
H\,{\scshape ii}\fi}
\newcommand{\mi}{\relax \ifmmode {\mu{\mbox m}}\else $\mu$m\fi}
\newcommand{\ha}{\relax \ifmmode {\mbox H}\alpha\else H$\alpha$\fi}
\newcommand{\hb}{\relax \ifmmode {\mbox H}\beta\else H$\beta$\fi}

\newcommand{\sii}{\relax \ifmmode {\mbox S\,{\scshape ii}}\else
S\,{\scshape ii}\fi}
\newcommand{\siii}{\relax \ifmmode {\mbox S\,{\scshape iii}}\else
S\,{\scshape iii}\fi}
\newcommand{\nii}{\relax \ifmmode {\mbox N\,{\scshape ii}}\else
N\,{\scshape ii}\fi}
\newcommand{\oi}{\relax \ifmmode {\mbox O\,{\scshape i}}\else
O\,{\scshape i}\fi}
\newcommand{\oii}{\relax \ifmmode {\mbox O\,{\scshape ii}}\else
O\,{\scshape ii}\fi}
\newcommand{\oiii}{\relax \ifmmode {\mbox O\,{\scshape iii}}\else
O\,{\scshape iii}\fi}
\newcommand{\neiii}{\relax \ifmmode {\mbox Ne\,{\scshape iii}}\else
Ne\,{\scshape iii}\fi}

\newcommand{\rdostres}{\relax \ifmmode {\,\mbox{R}}_{\rm 23}\else
\,\mbox{R}$_{\rm 23}$\fi}

\begin{document}

\selectlanguage{english}

\title{Not So Isolated: Green Pea Galaxies in Overdense Environments revealed by VLT/MUSE}

\selectlanguage{spanish}

\author{A. Arroyo-Polonio\inst{1}
        \and
        J.M. V\'{i}lchez\inst{2}
        \and
        J. Iglesias-P\'{a}ramo\inst{2},\inst{3}
        \and
        C. Kehrig\inst{2},\inst{4}
        \and
        E. P\'{e}rez-Montero\inst{2}
        \and
        R. Amor\'{i}n\inst{2}
        \and
        B. P\'{e}rez-D\'{i}az\inst{1}
        \and
        M. Hayes\inst{5}
        \and
        I. Breda\inst{6}
        \and
        J. S\'{a}nchez Almeida\inst{7},\inst{8}
        \and
        A. Gim\'{e}nez Alcazar\inst{2}
        \and
        M. Gonz\'{a}lez-Otero\inst{2}
        }

\institute{
            INAF–Osservatorio Astronomico di Roma, via di Frascati 33, I-00078 Monte Porzio Catone, Italy
            \email{antonio.arroyopolonio@inaf.it}
            \and
            Instituto de Astrof\'{i}sica de Andaluc\'{i}a, CSIC, Apartado de correos 3004, 18080 Granada, Spain 
            \and
            Centro Astron\'{o}mico Hispano en Andaluc\'{i}a, Observatorio de Calar Alto, Sierra de los Filabres, 04550 G\'{e}rgal, Spain
            \and
            Observat\'orio Nacional/MCTIC, R. Gen. Jos\'e Cristino, 77, 20921-400, Rio de Janeiro, Brazil
            \and
            Department of Astronomy, Stockholm University, AlbaNova University Centre, 106 91 Stockholm, Sweden
            \and
            Department of Astrophysics, University of Vienna, Türkenschanzstraße 17, 1180 Vienna, Austria
            \and
            Instituto de Astrof\'{i}sica de Canarias, E-38205 La Laguna, Tenerife, Spain 
            \and
            Departamento de Astrof\'{i}sica, Universidad de La Laguna, E-38206 La Laguna,Tenerife, Spain
            }

\selectlanguage{english}

\date{Received ?? / Accepted: ??}

\abstract
{

Context. Green Pea galaxies (GPs) are local starburst galaxies that serve as analogues for high-redshift star-forming galaxies, particularly Lyman continuum leakers. 
Historically considered isolated dwarfs, it remains debated whether their starbursts are driven by internal secular processes or external triggers.

Aims. We aim to constrain the role of the environment in this triggering. Specifically, we test whether external influence comes from close galaxy–galaxy interactions or more diffuse processes, such as gas accretion within overdense regions.

Methods. We analyse VLT/MUSE observations of 24 GPs at $z\sim0.2$, searching for galaxies with spectral line features to identify companions. We derive key physical properties (extinction, SFR, stellar mass, age, metallicity) for GPs and companions and estimate the dynamical mass of the groups.

Results. We identify 22 emission-line galaxies, 11 of which are companions ($|\Delta v| \leq 500$ km s$^{-1}$). We find a high companion fraction ($33^{+11}_{-8} \%$) and a $\sim$1 dex excess in number density compared to the field, confirming that GPs reside in overdense environments. However, companions typically lie at projected separations of $\sim$100 kpc with no evidence of ongoing interactions. Physically, while both populations show star-forming excitation, GPs form a homogeneous class of young (stellar mass-weighted age $\sim$230~Myr), metal-poor, high-sSFR starbursts with elevated velocity dispersions. In contrast, companions are more evolved ($\sim$1.6~Gyr) and heterogeneous, spanning broader ranges in stellar mass, metallicity, and dust attenuation. The inferred group dynamical masses are $\sim$3 dex higher than total stellar masses, suggesting significant dark matter and neutral gas content.

Conclusions. GPs do not appear to be triggered by ongoing major mergers with close (10–30 kpc) companions. Instead, results favor a scenario where GPs are transient starbursts in overdense regions, plausibly sustained by gas accretion. However, the limited spatial resolution prevents us from ruling out very close mergers ($\lesssim10$ kpc). Furthermore, the high dynamical-to-stellar mass ratios allow for substantial non-stellar mass components (i.e. dark matter and/or neutral gas) in these systems.

}

\keywords{}

\titlerunning{Starburst triggering}
\authorrunning{A. Arroyo-Polonio et al.}
\maketitle
\section{Introduction}
\label{seccion_intro}
Green Pea galaxies (GPs) are compact, low-mass ($M_\star \sim 10^9 \ M_\odot$) starbursts galaxies first detected at z $\sim$ 0.2, noted for their intense emission lines and high specific star formation rates (sSFR $\equiv \mathrm{SFR}/M_\star$) [$\sim(100 \ \mathrm{Myr})^{-1}$] \citep[e.g.,][]{cardamone2009galaxy,izotov2011green,arroyo2023muse}. They are considered nearby analogs of early-universe galaxies, sharing properties like low metallicity (12+log(O/H)  $\in$ (7.6, 8.4)), strong $[O\textsc{iii}]5007$ emission, and in some cases Lyman continuum (LyC) leakage \citep[e.g.,][]{amorin2010oxygen,izotov2016detection,izotov2016eight,yang2017lyalpha}. GPs represent a subset of the whole family of local extreme emission line galaxies (EELGs) \citep[e.g.][]{izotov2011green,perez2021extreme,breda2022characterisation,iglesias2022minijpas,lumbreras2022j,kouroumpatzakis2024blueberry,gimenez2025j}. What triggers these extreme starburst episodes in GPs remains an open question \citep[e.g.,][]{lofthouse2017local}. 

One longstanding hypothesis is galaxy interactions: in general, mergers and close encounters are known to induce starbursts, as classical studies have shown through disturbed morphologies and enhanced star formation in interacting galaxies \citep[e.g.,][]{joseph1985recent,kennicutt1987effects,barton2000tidally,ellison2008galaxy}. 
Indeed, the majority of local intense starbursts are associated with major mergers, which not only compress the gas and ignite powerful star formation, but also accelerate galaxy evolution and drive radial inflows that alter the global chemical composition \citep[e.g.,][]{sanders1996luminous,veilleux2002optical,perez2024departure}.
Star‐formation enhancement peaks in close pairs (those with projected separations $r_p\lesssim30$ kpc and line‑of‑sight velocity differences $|\Delta v|\lesssim300$ km s$^{-1}$) and declines rapidly at larger distances, implying that companions farther away are unlikely to trigger a central starburst \citep[][]{ellison2008galaxy}.

Whether GP starbursts are interaction‑induced is under active investigation. On one hand, individual cases of GPs do show clear signs of mergers \citep[e.g.,][]{purkayastha2022green,purkayastha2024second}. Here, HI imaging has uncovered close, gas‑rich companions and disturbed neutral‑gas reservoirs in a handful of systems, supporting merger‑triggered starbursts. On the other hand, a recent deep survey of 23 GPs with MUSE found no statistical excess of companions around GPs compared to normal star-forming galaxies
\citep{laufman2022triggering}.  Moreover, the companions identified in that survey tend to be at projected distances of 50–120 kpc, well beyond the range where interactions strongly enhance star formation \citep{ellison2008galaxy}. This led authors to conclude that these companions are probably unrelated to the starburst trigger in GPs. This conclusion is reinforced by statistical analyses of large EELG samples ($N>900$), which find low pair fractions ($f_{\mathrm{pair}} \sim 15 \%$) and suggest that internal processes, rather than external gravitational perturbations, are the primary drivers (M. Gonz\'{a}lez-Otero et al. in prep.). In other words, while mergers can produce starbursts, the current evidence suggests they may not be the dominant cause for most GPs. This situation mirrors the mixed results from studies of dwarf starbursts: e.g., luminous compact galaxies show a 36 $\%$ merger fraction \citep{rawat2007unravelling}, and blue compact dwarfs sometimes burst in isolation with no clear companion influence \citep{zitrin2009star}. Therefore, both merger-driven and merger-independent mechanisms must be considered when explaining the starburst trigger, especially the latter, which remains poorly understood and requires further investigation.

One possibility is that GPs arise in unusual environments or cosmic conditions that foster vigorous star formation. Rather than a single galaxy-galaxy collision, the trigger could be a broader overdensity of gas and dark matter on scales of tens or even hundreds of kiloparsecs \citep[see e.g.,][]{dekel2009cold}. In this scenario, a GP might be the most active member of a small group or filamentary structure, where multiple galaxies are simultaneously forming stars \citep[e.g.,][]{paudel2024discovery}. In such cases, the traditional view of an isolated starburst galaxy gives way to a picture of a starbursting region of the universe, a region possibly shaped by an underlying overdensity of dark matter and gas. 

An instant connection with this idea is cosmological gas accretion. Theoretical models predict that the cosmic web delivers cold gas into galaxies, and this process is especially effective for low-mass ($\lesssim10^{12}~M_\odot$) halos \citep{dekel2006galaxy}. Such pristine gas infall can spark new star formation without any interaction, essentially fueling a starburst from within. There are observational hints of this mechanism: many star-forming galaxies (including extremely metal-poor galaxies) exhibit kinematic irregularities in the gas and drops in metallicity best explained by recent accretion of metal-poor gas \citep{almeida2013local,ashley2014h,sanchez2014star,almeida2015localized,perez2024departure,billand2025investigating}.  In the context of GPs, this raises the intriguing idea that in some cases their starbursts might be triggered by a supply of fresh gas from their surroundings.
Several chemical and spectroscopic signatures indicate that many GPs were already chemically evolved before any accretion event took place. Their relatively high N/O ratios \citep[e.g.,][]{amorin2010oxygen,perez2011integral,sanchez2014star} and the detection of older stellar populations in their spectra \citep{amorin2012star} suggest a pre-existing enrichment phase. In this view, the accretion of metal-poor gas would not mark the onset of star formation, but rather a rejuvenation episode superimposed on an already evolved system.

In summary, GPs and their companions offer a fascinating laboratory to study extreme star formation and its drivers. The evidence to date points to a complex interplay of factors: although major mergers can trigger powerful bursts \citep[e.g.,][]{sanders1996luminous,veilleux2002optical}, most GPs show no close companions capable of igniting their star formation \citep[e.g.,][]{laufman2022triggering}. Instead, environmental and internal processes, like cold‑gas accretion in high-density \ion{H}{i} and dark matter halos, likely dominate. In this work, we aim to constrain these triggering mechanisms by characterising the immediate environment of a sample of 24 GPs using deep VLT/MUSE integral field spectroscopy. We conduct a systematic search for companions with spectral line features within the MUSE field of view and derive detailed physical properties (e.g., stellar masses, ages, metallicities, dynamical masses) for both the central starbursts and their neighbours to assess the state of these systems.

In Sect. \ref{seccion_observaciones} we describe the MUSE observations. Sect. \ref{seccion_companions} presents our methodology for detecting and confirming companions. In Sect. \ref{sec:phys_props} we derive the physical properties of both GPs and their companions. Sect. \ref{seccion_discussion} discusses the implications of these results for the mechanisms that may trigger the extreme star-formation episodes in GPs. Finally, Sect. \ref{seccion_conclusiones} summarises our main conclusions.
Throughout this paper, we use physical distances and assume a flat
cosmology with $H_0 = 70 \ \mathrm{km \ s^{-1} \ Mpc^{-1}}$, $\Omega_m = 0.3$, and $\Omega_\Lambda = 0.7$.

\section{Observations}
\label{seccion_observaciones}

We studied a sample of  24 GPs observed with MUSE \citep{bacon2010muse} at the Very Large Telescope (VLT; ESO Paranal Observatory, Chile). MUSE is a panoramic integral field spectrograph, which, operating in its wide field mode (WFM), provides a field of view (FoV) of $1' \times 1'$ with a spatial sampling of $0.2''$ and a full width half maximum (FWHM) spatial resolution of $0.3'' - 0.4 ''$.
The data were obtained in nominal mode (wavelength range $\lambda4750\AA-\lambda9350\AA$) with a spectral sampling of about $1.07 \ \mathrm{\AA pix^{-1}}$ and an average resolving power of $R \sim 3000$.
Our observational sample comprises the GPs from \cite{cardamone2009galaxy} at declinations < +20, ensuring visibility from the Paranal Observatory.
The program ID corresponding to the observations of these galaxies is $0102.B-0480(A)$ (PI: Hayes, Matthew).

We retrieved the fully reduced data cubes from the ESO archive. The data reduction was performed with MUSE Instrument Pipeline v. 1.6.1 with default parameters \citep{weilbacher2020data}, which consists of the standard procedures of bias subtraction, flat fielding, sky subtraction, wavelength calibration, and flux calibration.

This sample represents the overall properties of the GPs: e.g., the stellar masses, redshifts, metallicities, SFRs and [O\textsc{iii}]$\lambda$5007\AA \ EWs span the typical ranges reported for this class of objects (see the comparison between \cite{arroyo2023muse} and, e.g., \cite{cardamone2009galaxy,amorin2010oxygen}) .
The names, positions, redshifts, and information about the observations of the galaxies are in \cite{arroyo2023muse}.

\section{Searching for companions}
\label{seccion_companions}

Following the study by \citet{laufman2022triggering}, we revisit the search for companions around this sample of GPs. We find that $33^{+11}_{-8}\%$ of the GPs host at least one companion (within $\pm500$ km s$^{-1}$), restricting the definition of companion to $\pm300$ km s$^{-1}$ yields the same fraction.
The key to this new discovery lies in stacking the brightest optical emission lines, namely H$\alpha$, H$\beta$, [O\textsc{iii}]$\lambda$5007\AA, [O\textsc{iii}]$\lambda$4959\AA, and [O\textsc{ii}]~$\lambda\lambda$3726, 3729\AA\ (if present). 

To do so, we used the 24 MUSE datacubes covering the full FoV. For each GP, and based on its known redshift, we constructed a spectral mask for each line with a width of 250 km s$^{-1}$, as well as two continuum masks for each line group (i.e., H$\alpha$; H$\beta$ + [O\textsc{iii}]$\lambda$5007 + $\lambda$4959; and [O\textsc{ii}]$\lambda\lambda$3726, 3729). Each continuum mask was placed at 1500 km s$^{-1}$ away from the closest emission line within the group and had a width of 3000 km s$^{-1}$.

Using these masks, we created the so-called “nebular map” by stacking all line maps after continuum subtraction. In this construction,  nebular emission appears as a positive signal, while absorption lines produce negative signal. We therefore searched for both emission and absorption line systems using the same family of masks. Since companions may be offset in redshift/velocity relative to the GP, we applied velocity shifts to the masks in steps of 50 km s$^{-1}$, exploring up to $\pm$500 km s$^{-1}$ in both red and blue directions. We also extended the velocity range up to $\pm$5000 km s$^{-1}$ to search for unrelated  galaxies, reaching a total of 22 detections, all showing emission-line features (hereafter we can refer to them as ELGs). While these additional (11 more) galaxies are not classified as GP companions, their detection helps assess whether there is an overdensity of  galaxies near the GPs.

This approach is, however, insensitive to galaxies whose lines are absent or too weak to be detected. To quantify our sensitivity, we performed injection-recovery tests using artificial emission-line sources. We estimate a 90\% completeness limit for the stacked line flux ranging from $3.1 \times 10^{-17}$ erg s$^{-1}$ cm$^{-2}$ (in the best observing conditions) to $3.7 \times 10^{-16}$ erg s$^{-1}$ cm$^{-2}$ (in the worst). As an illustrative case, for a system at $z \sim 0.2$ with equal H$\alpha$ and [O\textsc{iii}] fluxes, our deepest limit corresponds to a minimum detectable SFR of $4 \times 10^{-3}$ M$_{\odot}$ yr$^{-1}$. Consequently, our search is inherently biased toward star-forming galaxies, meaning that passive or weakly star-forming companions falling below these thresholds may be missed. Objects with very low line equivalent widths, or with emission/absorption lines at low S/N, can remain undetected even if their stellar continuum is relatively bright. Therefore, our reported companion fraction should be explicitly regarded as a lower limit to the true galaxy population within the MUSE FoV.

Once the family of velocity-shifted nebular maps was generated, we identified candidate galaxies as those regions exhibiting significant signal above the background noise ($S/N > 3$). For each candidate, an integrated spectrum was extracted within the detected region. A visual inspection of these spectra was then performed to verify the presence of the expected emission lines (typically H$\alpha$, H$\beta$, and [O\textsc{iii}]), ensuring that spurious detections were excluded. Only sources with clear nebular-line signatures were confirmed as galaxies and considered companions when lying within $\pm500$ km s$^{-1}$ of the GP systemic velocity. After this point, to enhance the coherence among neighbouring spaxels, the datacubes were spatially smoothed using a Gaussian kernel with a standard deviation of two spaxels (corresponding to $\sim$0.4'' in MUSE data).

Regarding the kinematic analysis, velocity and velocity-dispersion maps were derived using the stacked emission-line profile, combining H$\alpha$, H$\beta$, [O\textsc{iii}]$\lambda$5007, and [O\textsc{iii}]$\lambda$4959.
For each spaxel that meets our signal-to-noise criterion ($S/N > 3$), a Gaussian fit is applied. We then measure the central wavelength and the standard deviation of the Gaussian that best fits the emission-line profile in each spaxel. The central wavelength of each emission line is crucial for determining the velocity field by using the longitudinal relativistic Doppler effect, expressed as:
\begin{equation}
v = c \left(\frac{(\lambda/\lambda_0)^2 - 1}{(\lambda/\lambda_0)^2 + 1}\right),
\end{equation}
where $v$ represents the velocity, $c$ the speed of light, $\lambda$ the observed wavelength, and $\lambda_0$ the rest wavelength.
This formula for the Doppler effect was consistently applied to all velocity calculations throughout our analysis. The Gaussian standard deviation provides the basis for calculating the velocity dispersion field.

To obtain the corrected velocity dispersion ($\sigma$), we accounted for both instrumental and thermal broadening ($\sigma_{\mathrm{inst}}$, $\sigma_{\mathrm{thermal}}$) as follows:
\begin{equation}
\sigma^2 = \sigma_{\mathrm{obs}}^2 - (\sigma_{\mathrm{inst}}^2 + \sigma_{\mathrm{thermal}}^2).
\end{equation}
The instrumental width was derived from the lamp calibration lines, yielding $\sigma_{\mathrm{inst}} = 0.9$ \AA, which at $\lambda \sim 5000$ \AA \ corresponds to an instrumental broadening of $\sigma_{\mathrm{inst}} \approx 54$ km s$^{-1}$.
The thermal width was estimated from the Maxwell–Boltzmann distribution:
\begin{equation}
\sigma_{\mathrm{thermal}}^2 = \frac{k T_e}{m},
\end{equation}
where $k$ is the Boltzmann constant, $T_e$ the electron temperature, and $m$ the hydrogen mass.
For a representative $T_e = 10^4$ K, this yields $\sigma_{\mathrm{thermal}} \approx 9$ km s$^{-1}$.
The correction varies only modestly across the expected temperature range of 8000–16 000 K ($\sigma_{\mathrm{thermal}} = 8$–12 km s$^{-1}$), so assuming $T_e = 10^4$ K provides a reliable approximation for our purposes.

\begin{figure}[h!]
\centering
   \includegraphics[width=\columnwidth]{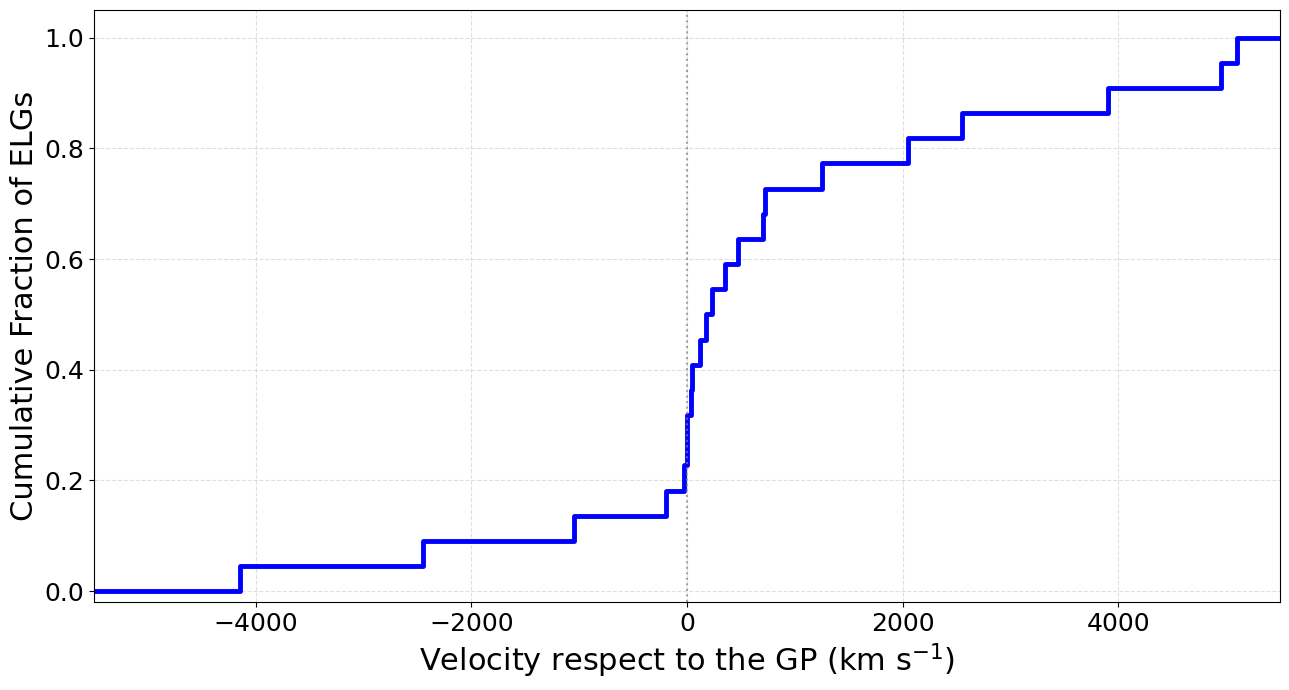}
     \caption{Empirical Cumulative Distribution Function of the velocity offsets relative to the GP systemic velocity for the 22 detected ELGs. The blue line represents the continuous distribution over the full searched range of $\pm$5000 km s$^{-1}$.}
     \label{histograma_velocidades_galaxias}
\end{figure}

\begin{figure*}[h!]
\centering
   \includegraphics[width=\textwidth]{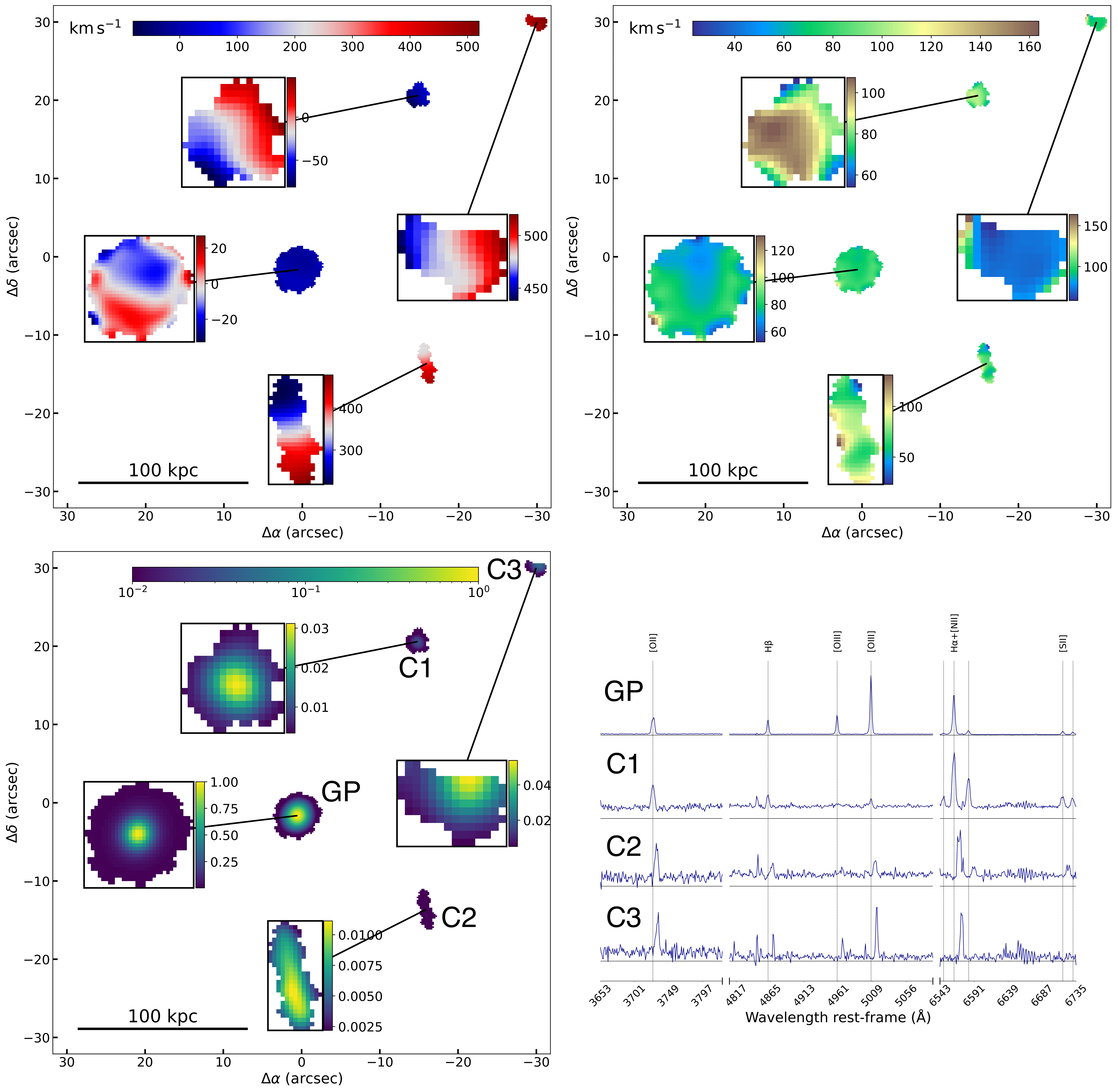}
     \caption{GP1 system. Each figure shows a cutout of each galaxy with its own colorbar. Spectra are sorted by the total flux of the sources, with the top spectrum corresponding to the brightest galaxy (always the GP). The top-left panel displays the velocity map, the top-right shows the velocity dispersion map, the bottom-left shows the flux map normalized to the maximum value, and the bottom-right presents the spectra of each galaxy, with the main optical emission lines marked at the rest frame of the GP. The figures in Appendix \ref{appendix} follow the same layout.}
     \label{panel_GP1}
\end{figure*}

Fig. \ref{histograma_velocidades_galaxias} shows the empirical cumulative distribution function of line-of-sight velocity offsets between ELGs and their associated GPs. The steepness of the distribution within $\pm500$ km s$^{-1}$ reveals a $\sim$1 dex excess in the ELG number density relative to larger offsets. Thus suggesting that the presence of a GP correlates with enhanced star formation not only within the galaxy itself but also in its immediate environment. Fig. \ref{panel_GP1} shows the full MUSE field of view around GP1, along with its three identified companions and their corresponding spectra, the rest of figures are in Appendix  \ref{appendix}.


\section{Physical properties of GPs and companions}
\label{sec:phys_props}


After identifying companion and background/foreground ELGs in the MUSE fields, we now characterise the physical properties of both the GPs and their companions. In this section we use four complementary tools: (i) optical nebular emission lines from the MUSE datacubes, which we use to derive dust extinction, star-formation rates (SFRs), ionisation diagnostics and kinematical properties; (ii) stellar population synthesis and spectral energy distribution (SED) fitting with Code Investigating GALaxy Emission (\textsc{cigale} v2019.2; \citealt{boquien2019cigale}), from which we obtain stellar masses and mass-weighted stellar ages; (iii) the \textsc{HII-CHI-mistry} code \citep{perez2014code}, which provides gas-phase abundances; and (iv) the dynamical mass estimator \citep{heisler1985estimating}, which allows us to infer the total dynamical mass of the galaxy groups from the projected positions and radial velocities of their members.



\subsection{Emission-line diagnostics: extinction, SFR, and BPT classification.}

For each GP and confirmed companion, we extract an integrated spectrum from the MUSE datacubes by co-adding all spaxels associated with the galaxy in the nebular maps (Sect.~\ref{seccion_companions}). From these spectra we measure the fluxes of the main optical emission lines, including \ha, \hb, [\oiii]$\lambda\lambda4959,5007$, [\oii]$\lambda\lambda3726,3729$, [\nii]$\lambda6584$, and [\sii]$\lambda\lambda6717,6731$ whenever detected with sufficient signal-to-noise ratio (S/N>3). Line fluxes are obtained by fitting single Gaussians after local continuum subtraction, and their uncertainties are derived using the bootstrap methodology \citep{efron1985bootstrap}.

\begin{figure}[h!]
\centering
   \includegraphics[width=\columnwidth]{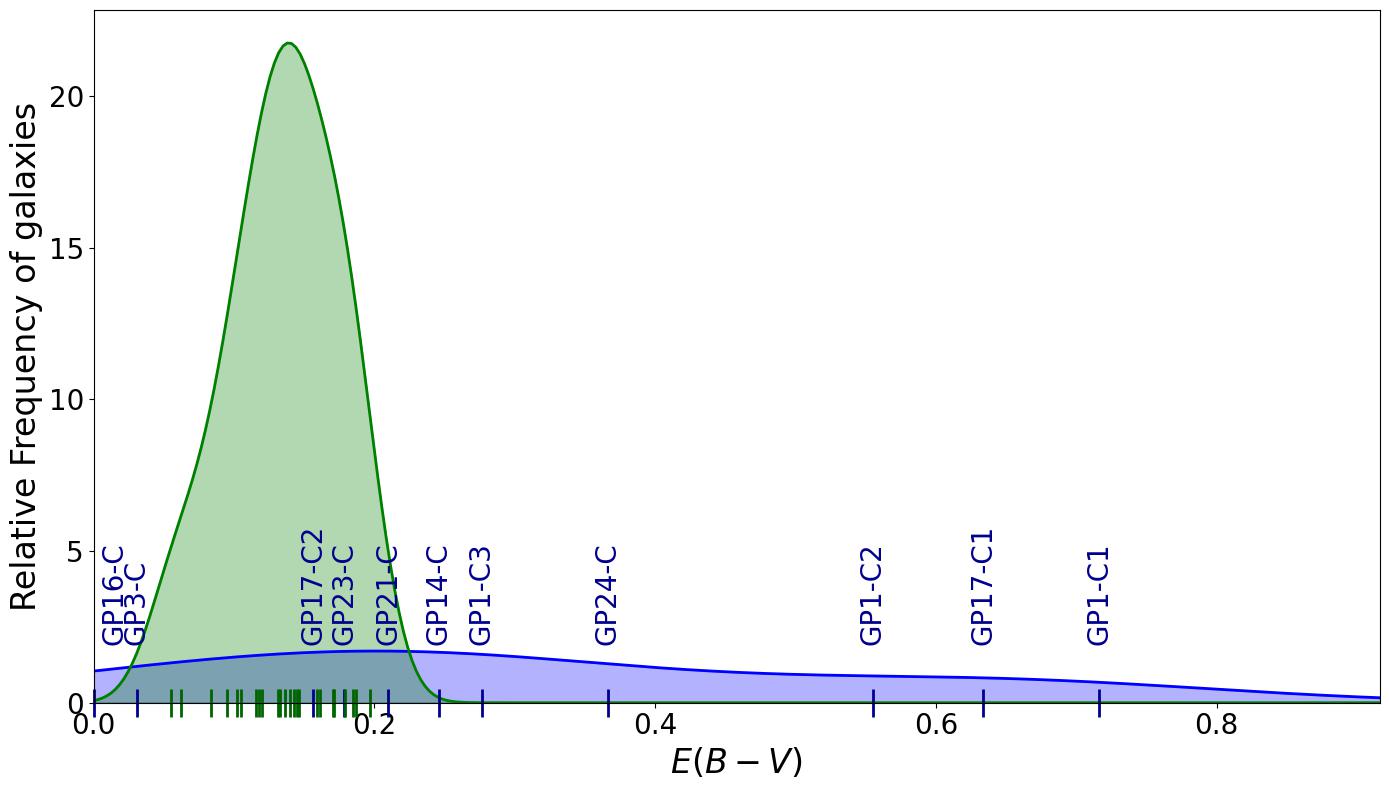}
     \caption{Distribution of the extinction $E(B-V)$ for GPs (green) and companion galaxies (blue), shown as a smoothed histogram. Individual galaxy positions and names are indicated by vertical marks and labels along the x\_axis.}
     \label{Extinction}
\end{figure}

Dust attenuation is derived from the Balmer decrement, using the observed \ha/\hb\ ratio compared to its Case~B recombination value for $T_e = 10^4$~K and $n_e = 100$~cm$^{-3}$ (intrinsic \ha/\hb\ = 2.86). To convert the Balmer excess into nebular colour excess $E(B-V)$, we adopt the extinction curve of \citet{cardelli1989relationship}, following the same approach commonly used for star-forming galaxies. The resulting distribution of $E(B-V)$ values for GPs and their companions is shown in Fig.~\ref{Extinction}.

We compute the intrinsic \ha\ luminosity, $L(\ha)$, from the extinction-corrected flux using the luminosity distance $D_L$ derived from the spectroscopic redshift of each galaxy, adopting the cosmological parameters defined in Sect.~\ref{seccion_intro} (i.e. $L = 4 \pi D_L^2 F_{\rm corr}$).
These luminosities are then used as our primary tracer of the current SFR. We convert \ha\ luminosities into SFRs assuming a continuous star-formation history and a Chabrier-like IMF, adopting the calibration
\begin{equation}
\mathrm{SFR} \ (\mathrm{M_\odot \ yr^{-1}}) = 5.5 \times 10^{-42} \ L(\ha) \ (\mathrm{erg \ s^{-1}}),
\label{equation}
\end{equation}
which is consistent with widely used updates of the original \ha--SFR relation \citep[e.g.][]{kennicutt1998global,kennicutt2012star}. In the following, we therefore treat SFR(\ha) as our reference indicator of the current star-formation rate for both GPs and companions.

\begin{figure}[h!]
\centering
   \includegraphics[width=\columnwidth]{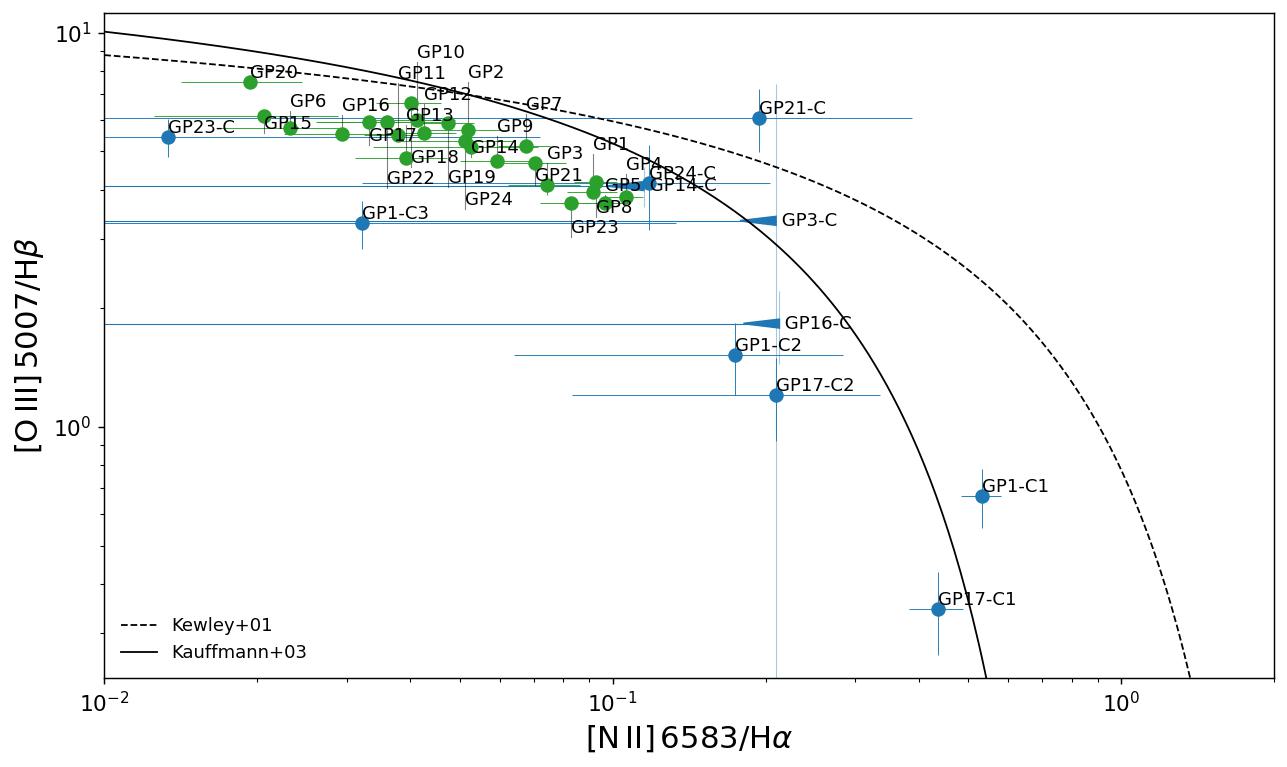}
     \caption{BPT diagram.  
              Green points represent the GPs, while blue points correspond to companion galaxies, labeled with the number of their associated GP.
              The lines that delimitate each region are taken from \cite{lineaBPT1} and \cite{lineaBPT2}.}
     \label{BPT}
\end{figure}

We use the classical Baldwin–Phillips–Terlevich (BPT) diagram \citep{baldwin1981classification,kewley2006host} to assess the dominant ionisation mechanism in both GPs and companions. For all galaxies we compute the line ratios 
$\log([\text{\oiii}]\lambda5007/\text{\hb})$ 
and 
$\log([\text{\nii}]\lambda6584/\text{\ha})$ 
whenever the four lines are reliably measured. In practice, all GPs and the majority of companions have \ha, \hb, [\oiii]$\lambda5007$, and [\nii]$\lambda6584$ detected with sufficient signal-to-noise to place them on the BPT plane. For three companions (GP3-C, GP14-C, and GP16-C) [\nii]$\lambda6584$ is not detected; for these objects we estimate an upper limit to the [\nii] flux from the rms of the local continuum and use this as an upper limit on $\log([\text{\nii}]/\text{\ha})$ in the diagram. The resulting distribution of GPs and companions in the BPT plane is shown in Fig.~\ref{BPT}.

\begin{figure}[h!]
\centering
   \includegraphics[width=\columnwidth]{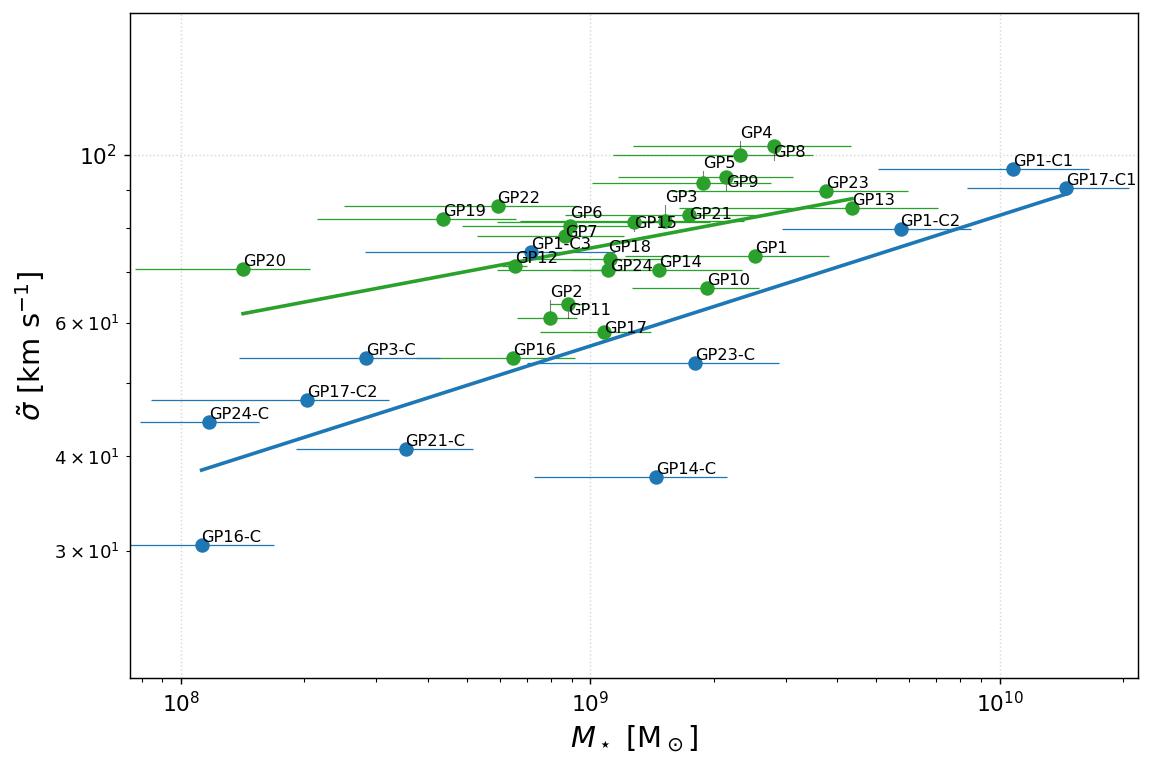}
     \caption{Flux-weighted velocity dispersion vs stellar mass.  
              Green points represent the GPs, while blue points correspond to companion galaxies.
              The lines in the same color are the lineal regression of the two groups of galaxies respectively.}
     \label{sigma}
\end{figure}

Finally, we derive the flux-weighted velocity dispersion,
\begin{equation}
\tilde{\sigma} \equiv \frac{\sum_i F_i \, \sigma_i}{\sum_i F_i},
\end{equation}
using the nebular flux and velocity-dispersion maps (see, e.g., Fig.~\ref{panel_GP1}). We then compare $\tilde{\sigma}$ with the stellar mass $M_\star$ derived in the next section. Fig.~\ref{sigma} shows the correlation between these two physical parameters for both the GPs and their companion galaxies.

\subsection{Stellar masses and mass-weighted stellar ages from \textsc{cigale}.}

We use \textsc{cigale} to derive stellar masses and mass-weighted stellar ages for all GPs and companions. \textsc{cigale} builds a large grid of synthetic SEDs based on parametric star-formation histories, stellar population models, nebular emission, and dust attenuation, and compares them to the observed photometry. For each galaxy, the code computes a $\chi^2$ value for every model and constructs probability distribution functions (PDFs) for the physical parameters using a Bayesian-like approach. In this work we adopt the median of each PDF as our fiducial estimate and the 16th–84th percentile range as its formal uncertainty.

The photometric input to \textsc{cigale} consists of optical ``photospectra'' constructed from the MUSE datacubes. We convolve each galaxy spectra with a set of synthetic narrow-band filters that densely sample the optical range. This yields a set of $\sim$50 narrow-band fluxes per object, providing a sampling of the rest-frame optical SED from $\sim$3800 to $\sim$9000~\AA. The SED fits are driven entirely by the optical continuum and the contribution of nebular emission. The redshift of each galaxy is fixed to its spectroscopic value.

The star-formation history is modelled with a two-component parameterisation (\texttt{sfhdelayed}) that captures both the underlying stellar population and the recent starburst. We use the \cite{bruzual2003stellar} stellar population models with a \cite{chabrier2003galactic} IMF. The parameter space explored is summarized in Table~\ref{tab:cigale_params}. It includes a wide grid of ages and e-folding times for both the main population and the burst, as well as various stellar metallicities ($Z$). For the nebular component, we varied the ionisation parameter ($\log U$) and the gas metallicity ($Z_{gas}$). Dust attenuation was implemented following a \cite{calzetti2000dust} law, allowing for different extinction levels for the young and old stellar populations. In our analysis, we utilize the \textsc{cigale} SED fits to derive the total stellar mass $M_\star$ and the mass-weighted stellar age $\langle t_\star \rangle_{\rm M}$.

\begin{table}

\centering
\begin{tabular}{ll}
\toprule
Parameter & Values \\
\midrule

$\tau_{\text{main}}$ (Myr) & 100, 300, 1000, 3000 \\
$age_{\text{main}}$ (Myr) & 100, 200, 500, 1000, 2000, 5000 \\
$\tau_{\text{burst}}$ (Myr) & 2, 5, 10, 30, 100 \\
$age_{\text{burst}}$ (Myr) & 5, 30, 80 \\
$f_{\text{burst}}$  & 0.01, 0.03, 0.05, 0.1, 0.2, 0.4, 0.6 \\

Metallicity $Z$ & 0.0001, 0.0004, 0.004, 0.008, 0.02 \\

$\log U$  & $-$3.7, $-$3.2, $-$3.0, $-$2.8, $-$2.6, $-$2.4, \\
 & $-$2.0, $-$1.7 \\
$Z_{\text{gas}}$ & 0.002, 0.004, 0.008, 0.016, 0.025 \\
$n_e$ (cm$^{-3}$) & 100 \\

$E(B-V)_{\text{young}}$ & 0.2, 0.3, 0.4, 0.5 \\
$E(B-V)_{\text{old\_factor}}$ & 0.3, 0.5, 1.0 \\
\bottomrule
\end{tabular}
\caption{Parameter grid used for the \textsc{cigale} SED fitting.}
\label{tab:cigale_params}
\end{table}



\begin{figure}[h!]
\centering
   \includegraphics[width=\columnwidth]{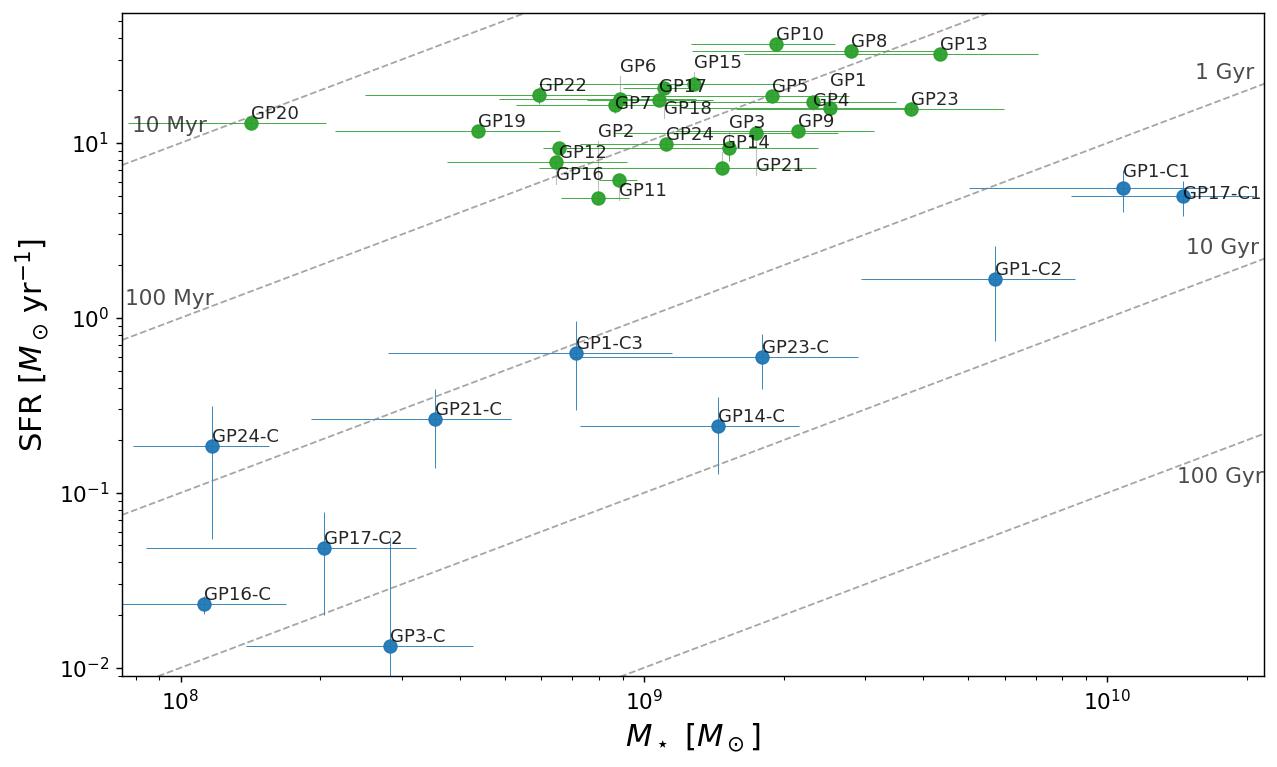}
     \caption{SFR vs stellar mass for GPs (in green) and their companion galaxies (in blue). Dashed lines represent constant sSFR, labeled by their corresponding mass doubling times (1/sSFR).}
     \label{SFR_stellar_mass}
\end{figure}

Figure~\ref{SFR_stellar_mass} shows the distribution of GPs and companions in the SFR–$M_\star$ plane, using SFR(\ha) as defined in Eq. \ref{equation} and stellar masses from \textsc{cigale}. A detailed discussion of their location relative to the local star-forming main sequence is deferred to Sect.~\ref{seccion_conclusiones}.

Given the known limitations of optical-only SED fitting, we report the median mass-weighted stellar ages from \textsc{cigale} for each group (GPs and companions) rather than individual-galaxy values.
For each group we characterise the age distribution using the median and the 16th–84th percentile range, a non-parametric interval equivalent to a $1\sigma$ uncertainty for a Gaussian distribution.
For the GPs, the mass-weighted stellar age is $\sim$236~Myr with a $+146/-94$~Myr interval.
For the companions, the corresponding values are $\sim$1.60~Gyr with a $+256/-750$~Myr interval.



\subsection{Gas-phase abundances from \textsc{HII-CHI-mistry}.}

We use the optical version 6.0 of the \textsc{HII-CHI-mistry} code (\textsc{HCm}; \citealt{perez2014code}) to derive gas-phase oxygen abundances of the companions. We specifically adopted the model grids calculated using the POPSTAR stellar synthesis libraries \citep{molla2009popstar}, as they provide the most appropriate description of the ionising radiation field for regular star-forming galaxies. \textsc{HCm} compares observed strong-line ratios with these grids of photoionisation models that span wide ranges in 12+log(O/H), and returns model-based estimates that are consistent, when [\oiii]$\lambda4363$ is available, with the direct $T_e$ method. For each galaxy the code computes a likelihood for every model in the grid and constructs probability distribution functions for the metallicity. As in the case of \textsc{cigale}, we adopt the median of each PDF as our fiducial value of 12+log(O/H) and the 16th–84th percentile range as the associated uncertainty.

The input line fluxes for \textsc{HCm} are measured from the same integrated spectra used in the emission-line analysis described above. We consider all available optical strong lines with sufficient signal-to-noise, typically including [\oii]$\lambda\lambda3726,3729$, \hb, [\oiii]$\lambda\lambda4959,5007$, \ha, [\nii]$\lambda6584$, and [\sii]$\lambda\lambda6717,6731$. In practice, we restrict the \textsc{HCm} analysis to galaxies with a reliable measurement or constraint on [\nii]$\lambda6584$, which is required to break the degeneracy between O/H and N/O in the model grid. This includes 
all but three companions (GP3-C, GP14-C, and GP16-C). For these three objects [\nii] is too weak to be detected; they are therefore excluded from the \textsc{HCm}-based abundance analysis. For the remaining galaxies, \textsc{HCm} uses the subset of line ratios available in each case, following its internal prescriptions for combining information from different diagnostics.

\begin{figure}[h!]
\centering
   \includegraphics[width=\columnwidth]{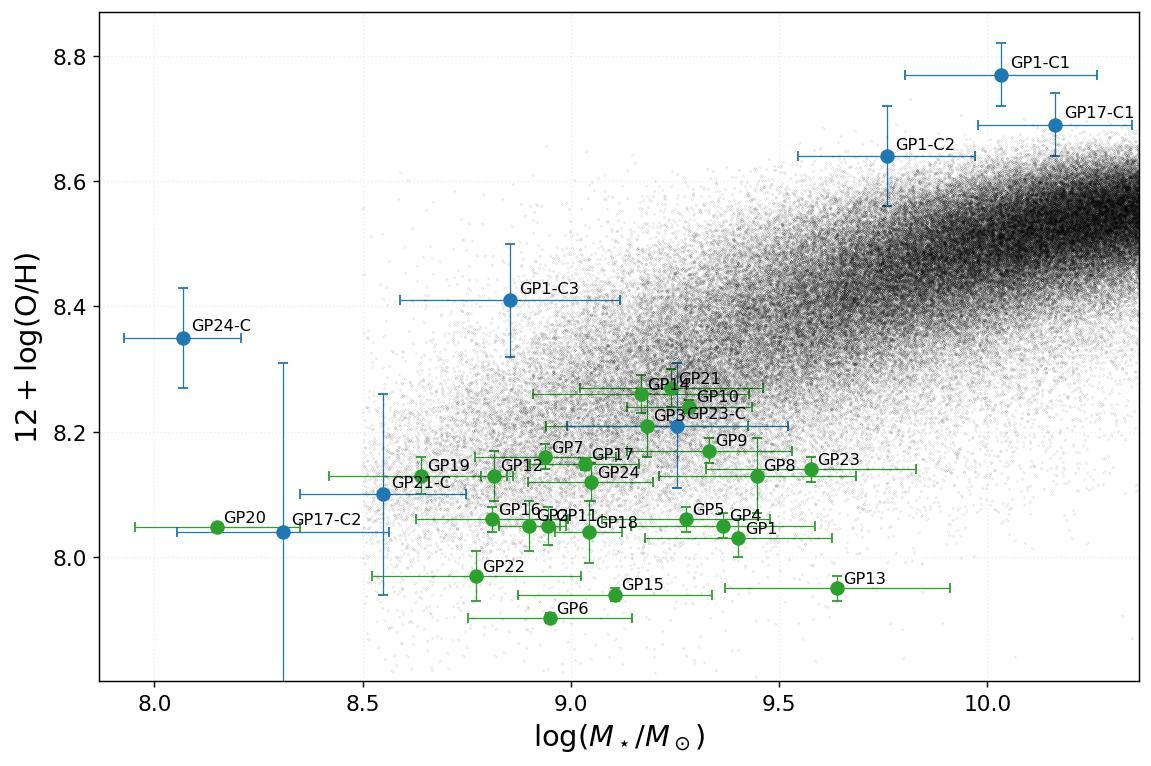}
     \caption{Metallicity vs stellar mass for GPs (in green) and their companion galaxies (in blue). Black points correspond to $\sim$200 000 galaxies from
\cite{puertas2022mass}.}
     \label{HCm_Z_vs_Mstar}
\end{figure}

The 12+log(O/H)  values for GPs are taken from \citet{arroyo2023muse}, where the detection of the [\oiii]$\lambda4363$ line in some GPs enables direct $T_e$ abundances, providing more reliable estimates.
The resulting oxygen abundances  are combined with the stellar masses derived from \textsc{cigale} to build a set of standard scaling relations for GPs and their companions. Figure~\ref{HCm_Z_vs_Mstar} shows the distribution of 12+log(O/H) as a function of stellar mass. 

\subsection{Dynamical mass of the galaxy groups}
\label{sec:dynamical_mass}

To investigate the gravitational potential in which these systems reside, we estimated the total dynamical mass ($M_{\mathrm{dyn}}$) of the galaxy groups identified in our sample. We applied the dynamical mass estimator proposed by \citet{heisler1985estimating}, which is suitable for groups with a small number of members. The total dynamical mass is given by:

\begin{equation}
M_{\mathrm{dyn}} = \frac{32}{\pi} \frac{1}{G (N-3/2)} \sum_{i}^{N} r_{i} v_i^2,
\end{equation}

where $v_i$ is the radial velocity of the $i$-th galaxy relative to the system's center of mass, $r_{i}$ is its projected distance from the center, $G$ is the gravitational constant, and $N$ is the total number of group members. The center of mass position and velocity were determined by weighting each member galaxy by its individual stellar mass, as previously estimated. We restricted this calculation to systems with at least three detected galaxies (the GP and a minimum of two companions) to ensure a minimal statistical stability. Given the small number of members, these estimates are subject to large statistical uncertainties and projection effects. To quantify these, we performed a Monte Carlo analysis for each system. In each iteration, we perturbed the positions ($x, y$) and radial velocities ($v_i$) using a Gaussian distribution based on the observed group dispersions ($\sigma_{\mathrm{pos}}$ and $\sigma_{v}$). We adopted the 16th and 84th percentiles of the resulting mass distribution as the $1\sigma$ confidence interval.

Applying this estimator to the group hosting GP1 ($N=4$), we derive a dynamical mass of $M_{\mathrm{dyn}} = 2.64 \times 10^{13} \ M_\odot \ (+0.55 / -0.24 \ \mathrm{dex})$. In comparison, the total stellar mass of the system (summing the GP and their companions) is $M_{\star} = 2.0 \times 10^{10} \ M_\odot$, implying a dynamical-to-stellar mass ratio of $\sim 1300$. Similarly, for the GP17 group ($N=3$), we obtain $M_{\mathrm{dyn}} = 1.19 \times 10^{13} \ M_\odot \ (+0.51 / -0.56 \ \mathrm{dex})$ versus a total stellar mass of $1.6 \times 10^{10} \ M_\odot$.  These ratios are higher than those found in recent studies of similar systems; for instance, \citet{paudel2024discovery} reported dynamical masses typically two orders of magnitude larger than the stellar component in dwarf galaxy groups.

To mitigate small-number statistics in individual groups, we performed a stacking analysis using the 24 GP systems. We centered the reference frame on each GP and stacked the phase-space coordinates of all identified companions. Using the full sample of companions, we obtain a dynamical mass of $2.6 \times 10^{12} \, M_\odot  \ (+0.29 / -0.13 \ \mathrm{dex})$ compared to an average stellar mass of $3.0 \times 10^{9} \, M_\odot$, recovering the $\sim 3$ dex difference observed in individual cases. However, if we restrict the calculation to companions within a tighter velocity window of $\pm 300 \, \mathrm{km \, s^{-1}}$ (to exclude potential high-velocity interlopers or unbound flybys), the dynamical mass decreases to $7.1 \times 10^{11} \, M_\odot  \ (+0.31 / -0.1 \ \mathrm{dex})$, while the average stellar mass remains similar ($2.8 \times 10^{9} \, M_\odot$). This results in a dynamical-to-stellar mass ratio of $\sim 250$, which is consistent with the factor of $\sim 100$ reported by \citet{paudel2024discovery}.

\section{Discussion}
\label{seccion_discussion}

The main goal of this work is to investigate whether the extreme starbursts observed in GPs are primarily driven by direct galaxy–galaxy interactions, or instead by more diffuse environmental processes such as gas accretion within overdense regions. Our analysis combines a census of emission-line companions around 24 GPs with a detailed characterisation of their stellar and gas-phase properties. In this section we discuss the implications of our results for the triggering of the GP starbursts.

\subsection{Environmental overdensity around GPs}

A first key result is the overdensity of ELGs in the vicinity of GPs in velocity space. As shown in the Empirical Cumulative Distribution Function of Fig.~\ref{histograma_velocidades_galaxias}, the distribution exhibits a prominent steepening within $\pm500$~km~s$^{-1}$ of the GP systemic velocity, showing an excess of nearly 1~dex relative to larger velocity offsets. We identify $11^{+4.4}_{-3.3}$ companions within this range (all uncertainties represent 68$\%$ confidence intervals for Poisson statistics following \cite{gehrels1986confidence}) and a further $11^{+4.4}_{-3.3}$ ELGs at larger offsets (up to $\pm5000$~km~s$^{-1}$).

To quantify this environmental enhancement, we performed a number density estimation assuming a standard cosmology at the median redshift of the sample ($z \sim 0.25$). The 24 MUSE pointings cover a total volume of $\sim 170$~Mpc$^3$, spanning the velocity range of $\pm 5000$~km~s$^{-1}$. Using the 11 field galaxies detected in the outer velocity windows ($|\Delta v| > 500$~km~s$^{-1}$), we derived a background galaxy density of $\sim 0.07^{+0.03}_{-0.02}$~galaxies~Mpc$^{-3}$.

Applying this background rate to the total volume spanned by the companion windows ($\sim 17$~Mpc$^3$), we estimate an expected background of $\lambda = 1.2^{+0.5}_{-0.4}$ interlopers. Assuming a Poisson distribution, the probability of observing 11  galaxies against this background expectation is highly significant ($p < 10^{-7}$). This leaves a net count of $9.8^{+4.4}_{-3.3}$ physically associated galaxies, corresponding to a physical number density of $0.6^{+0.26}_{-0.20}$~galaxies~Mpc$^{-3}$ for the GP companions. Consequently, even after correcting for expected background contamination, the local environment of GPs exhibits a density of galaxies $8.0^{+4.6}_{-3.1}\,(1\sigma)\,^{+11.8}_{-5.1}\,(2\sigma)\,^{+23.2}_{-6.5}\,(3\sigma)$ times higher than the surrounding field derived from the same dataset.   

Although the $\pm500$~km~s$^{-1}$ threshold is broader than the 200-300~km~s$^{-1}$ range typically used to define gravitationally bound pairs, it effectively characterizes the larger-scale environmental overdensity. To address possible contamination from projection effects or unbound flybys, we evaluated tighter velocity windows. Within a range of $\pm300$~km~s$^{-1}$, we detect 9 companions against a predicted Poisson background of $\lambda = 0.7^{+0.3}_{-0.2}$ ($p < 10^{-8}$). Even at $\pm200$~km~s$^{-1}$, we find 8 galaxies where only $\sim$0.5 are expected by chance. In terms of incidence, $33^{+11}_{-8}\%$ of the GPs host at least one companion within $\pm500$~km~s$^{-1}$, and this fraction remains unchanged when adopting these tighter cuts (even at the $\pm200$~km~s$^{-1}$ window). These results demonstrate that the observed excess is a robust physical feature and not an artifact of projection effects or velocity threshold selection. Together, these results indicate that GPs tend to reside in environments where star-forming galaxies are more common than in the average field, consistent with small groups or filamentary structures rather than truly isolated dwarfs. At the same time, the projected separations of most companions are typically $\gtrsim100$~kpc, with the closest system at $\sim30$~kpc (GP24). These separations lie largely outside the regime where the strongest interaction-driven star-formation enhancements are expected in close pairs \citep[e.g.,][]{ellison2008galaxy}, suggesting that the observed overdensity primarily traces the large-scale environment rather than ongoing major mergers in most cases.

Our finding of a localized overdensity on scales of $\sim100$~kpc must be contextualized within the broader framework of previous environmental studies. Historically, studies have shown that compact star-forming systems, such as HII galaxies, tend to reside in globally low-density environments compared to normal field galaxies \citep[e.g.,][]{campos1993empirical,vilchez1995spectroscopic,telles2000local}. Recent large-scale spatial distribution analyses confirm this profound isolation for GPs on megaparsec scales \citep[e.g., using a 5~Mpc radius;][]{gupta2026blueberry}. However, this macro-scale isolation does not preclude the existence of small-scale substructures. In fact, our detection of a high fraction ($33^{+11}_{-8}\%$) of local emission-line companions directly matches the findings of \cite{noeske2001faint}, who found faint dwarf companion candidates for $\sim30\%$ of star-forming dwarf galaxies within remarkably similar limits (projected separation $<100$~kpc and $\Delta v < 500$~km~s$^{-1}$). Similarly, \cite{pustilnik2001environment} demonstrated that when deeper searches are conducted, a large fraction of blue compact galaxies possess nearby companions. Our deep MUSE observations focus on the small ($\lesssim$ 200 kpc) scales. While GPs avoid massive cosmic structures, they frequently inhabit small, locally overdense pockets consisting of faint, gas-rich companions. A population that is systematically missed in large-scale or shallow surveys.

\subsection{Characterizing the GPs and their companions}

The emission-line properties help to clarify the nature of both GPs and their companions. The BPT diagram (Fig.~\ref{BPT}) shows that both GPs and companions are overwhelmingly consistent with being dominated by stellar photoionisation. The GPs occupy a compact region on the high-excitation/low metallicity star-forming branch, reflecting their uniform, extreme ionising conditions. In contrast, the companions are more broadly distributed across the star-forming sequence, including one object (GP1-C1) in the composite region and another (GP21-C) formally in the AGN area of the diagram. These isolated cases suggest that AGN-like ionisation may occur in a minority of companions, but they remain exceptions rather than the rule. 
The extinction distribution (Fig.~\ref{Extinction}) reveals a broader range of $E(B-V)$ among companions, including three heavily obscured systems with $E(B-V)>0.5$ (GP1-C1, GP17-C1 and GP1-C2). These three objects also stand out as the most massive companions in the sample, reaching in some cases stellar masses greater than $\sim10^{10}\,M_\odot$ and exceeding the mass of all GPs in the sample. In contrast, most GPs exhibit more moderate attenuation ($E(B-V)<0.2$). In addition, the $\tilde{\sigma}$-$M_\star$ plane (Fig.~\ref{sigma}) shows a clear correlation when both quantities are expressed on a logarithmic scale, albeit with substantial scatter. Such an approximately power-law behaviour is expected for virialised systems (since $M \propto R\,\sigma^2$), and the scatter is naturally enhanced here because we use stellar (rather than dynamical or total) masses, the characteristic radius $R$ varies from galaxy to galaxy (and is not constrained in our analysis), and the ionised gas in these star-forming systems is not required to be in virial equilibrium. We also find that GPs systematically show higher $\tilde{\sigma}$ than their companions at fixed $M_\star$, pointing to an excess of kinetic energy in the ionised gas that is plausibly linked to the intense stellar feedback associated with their compact starburst activity.
 More generally, companions span a much wider range of observed properties (extinction, excitation, kinematics) while GPs form a much more homogeneous class of compact, extreme starbursts within this broader environmental population.

The stellar population analysis based on \textsc{cigale} highlights a marked difference between GPs and their companions. GPs have very young mass-weighted stellar ages, with a median of $\sim$240~Myr and a relatively narrow, though asymmetric, $+150/-90$~Myr scatter. This is consistent with a dominant young stellar component that has formed on short timescales, in line with their extreme sSFRs. In contrast, the companions exhibit substantially older and more heterogeneous stellar populations, with a median mass-weighted stellar age of $\sim$1.60~Gyr and a broad, asymmetric $+260/-750$~Myr interval. These age differences indicate that, even though both GPs and companions reside in the same large-scale environment, the current starburst episode is strongly preferentially hosted by the GPs. The positions of the two populations in the SFR–$M_\star$ plane (Fig.~\ref{SFR_stellar_mass}) reinforce this picture: GPs populate the high-sSFR envelope of low-mass extreme SF galaxies, while most companions lie close to, the locus expected for ``normal'' local SF galaxies \citep[e.g.,][]{whitaker2012star}.

The comparison of stellar masses between GPs and companions adds an important constraint. Companions span a wider range in $M_\star$ than the GPs, reflecting the general fact that they also cover a broader range of physical properties. In two systems, at least one companion is more massive than the GP itself (e.g. GP1, which hosts two more massive companions, and GP17), demonstrating that companions are not necessarily low-mass satellites of a dominant central galaxy. In other two cases (such as GP14 and GP23), companions have stellar masses of the same order as their associated GP, while in the remaining four systems the GP is typically about an order of magnitude more massive than its companions. Taken together, these results show that GPs inhabit environments where they can be either the most massive, a peer, or even the less massive member of a small-scale system. The extreme starburst phase therefore does not seem to be tied to a specific mass hierarchy within the local group.

The gas-phase abundances derived with \textsc{HCm} add an important chemical dimension to this comparison. In the mass-metallicity plane (Fig.~\ref{HCm_Z_vs_Mstar}), the GPs populate the low-mass, low-metallicity regime characteristic of EELGs. The companions, however, span a much broader range in both stellar mass and metallicity. Among the subset of companions with stellar masses comparable to the GPs, the trend toward higher metallicity at fixed mass is only mild: two systems lie noticeably above the GP metallicities, while the remaining ones overlap with the GP locus within uncertainties. In contrast, the three most massive companions reach metallicities above $12+\log(\mathrm{O/H})\gtrsim8.6$, clearly separating them from the GPs. Overall, this diversity reflects a chemically heterogeneous companion population, within which some galaxies appear more evolved than the GPs while others share similar metallicities and comparable stellar masses.

\subsection{Particular companions}

There are two companions (GP1-C2 and GP17-C1) that stand out due to their remarkably high rotation velocities. These galaxies exhibit double-peaked emission lines (see Figs.~\ref{panel_GP1} and \ref{panel_GP17}), with peak separations of $\sim$ 200 km s$^{-1}$. They also share similar properties with GP1-C1, being the three companions with higher stellar mass, metallicity and gas attenuation. 

The three companions most similar to the GPs are GP23-C, GP24-C, and GP1-C3. These galaxies have stellar masses comparable to those of the GPs (except for GP24-C, being 1 dex less massive). They occupy nearly the same locus in the BPT diagram. The main difference is their specific star-formation rate: although these companions lie among the objects with the highest sSFR in the companion sample, their sSFRs still remain roughly one dex below those of the GPs.

In addition, GP17 shows a southern protrusion in the nebular maps (Fig.~\ref{panel_GP17}) whose ionised gas is redshifted by $\sim$50~km~s$^{-1}$ with respect to the bulk of the GP. This feature also displays a lower velocity dispersion, by $\sim$20~km~s$^{-1}$ compared to the GP average. If interpreted as a distinct system, it would correspond to a very close companion at a projected distance of $\sim$15~kpc from the GP centre, making it a plausible ongoing merger candidate. However, given the uncertainty on whether this structure is truly a separate galaxy and our inability to constrain its line-of-sight separation, we cannot confirm this scenario.

\subsection{A unified scenario for starburst triggering}

Putting all these pieces together, our results suggest a picture in which the extreme starbursts in GPs are not primarily triggered by ongoing major mergers with close (10-30 kpc) companions. Instead, GPs appear to sit in overdense regions populated by a heterogeneous mix of evolved and less extreme star-forming galaxies. 

The dynamical properties of these environments provide compelling support for this scenario. The dynamical masses estimated for the GP groups are typically $\sim$2.5-3 dex higher than their total stellar masses. Even when restricting the analysis to the closest companions in velocity space ($\pm 300$ km s$^{-1}$), the dynamical-to-stellar mass ratio remains high ($\sim 250$). This substantial discrepancy implies that the visible galaxies are possibly embedded in a deep gravitational potential dominated by a massive dark matter halo and, likely, a rich reservoir of neutral gas in the intragroup medium.
Furthermore, the significant drop in dynamical mass when tightening the velocity cuts suggests that a fraction of the identified companions may be unbound flybys rather than virialised members. Consequently, we caution that the true dynamical mass could be lower than estimated if these galaxies are not gravitationally bound, as the inclusion of high-velocity flybys in virial estimators inevitably leads to an overestimation of the total mass.

In this framework, a plausible driver of the GP starbursts is the rapid accretion of metal-poor gas (likely of predominantly pristine origin) from an IGM that, in some cases, could be already moderately enriched before the current burst.
The high dynamical-to-stellar mass ratios and the observed overdensity of ELGs within $\pm 500$~km~s$^{-1}$ are consistent with a shared large-scale reservoir (anchored by the local dark-matter overdensity) which acts as the fuel source. Whether in the form of group-scale HI clouds or filamentary inflows, this reservoir is capable of sustaining the extremely high sSFRs observed in GPs. The companions, by contrast, trace the more ``normal'' star-forming response to this same dense environment: while they may experience enhanced activity, they lack the specific conditions (such as a central position in the potential well or a recent rejuvenation event) to be pushed into the extreme bursting regime characteristic of the GPs.

\subsection{Limitations and caveats}

An important additional limitation of our study is the spatial resolution. At the redshifts of our sample ($z\sim0.2$), the typical MUSE point spread function corresponds to physical scales of $\sim$5--15~kpc \citep[e.g.][]{arroyo2023muse}, so our data are insensitive to very close companions or merger signatures on smaller scales. By contrast, the merger scenarios around nearby ($z<0.05$) GPs studied by \citet{purkayastha2022green,purkayastha2024second} involve companions at projected separations of only $\sim$4.7~kpc and $\sim$17.5~kpc. Systems like these would be only marginally resolved, or entirely unresolved, in our data. We therefore cannot rule out that our GPs are undergoing very close interactions that remain hidden in our seeing-limited observations.
The abundance estimates for companions rely on strong-line methods and on the availability of [\nii] for most galaxies. Furthermore, projection effects can complicate the association of companions in velocity space with the true three-dimensional environment.

\section{Summary and conclusions}
\label{seccion_conclusiones}

In this work we have used VLT/MUSE observations of 24 GPs at $z\sim0.2$ to revisit the role of environment in triggering their extreme starbursts. We combined a systematic search for line features in the full MUSE field of view with a homogeneous characterisation of the stellar, dynamical, and gas-phase properties of both GPs and their companions, using optical emission lines, \textsc{cigale} SED fitting, \textsc{HCm}, and dynamical mass estimators.

Our results suggest that GPs reside in overdense, star-forming environments, at least as traced by the population of ELGs. We identify 22 ELGs in the MUSE fields, 11 of which lie within $\pm 500$~km~s$^{-1}$ of the GP systemic velocity and are classified as companions. We find that $33^{+11}_{-8}\%$ of the GPs host at least one companion within this range. The empirical cumulative distribution of line-of-sight velocity offsets between ELGs and their associated GPs shows a $\sim$1 dex excess at $\pm500$ km s$^{-1}$ relative to larger offsets, providing a compelling, albeit limited by sample size, indication that GPs are embedded in small-scale structures where star formation is enhanced not only in the GP itself but also in its surroundings. Companions occupy a much broader range of observed properties (extinction, excitation, chemical abundances, stellar mass, stellar age) than the GPs, which appear as a more homogeneous population of compact, high-excitation starbursts.


Stellar masses and ages derived with \textsc{cigale} reveal that the starburst phase is strongly concentrated in the GPs. GPs have very young mass-weighted stellar ages, with a median of $\sim$230~Myr, while companions are significantly older, with a median of $\sim$1.6~Gyr and a much larger intrinsic scatter. In the SFR–$M_\star$ plane, GPs lie on the high-SFR envelope of low-mass star-forming galaxies, whereas most companions are closer to the locus of ``normal'' SF galaxies. The relative mass hierarchy varies from system to system: in some GPs (e.g. GP1, GP17) at least one companion is more massive than the GP; in others (GP14, GP23) the masses are comparable, and in the remaining cases the GP is typically about one order of magnitude more massive than its companions. Thus, the occurrence of the extreme starburst does not correlate with a fixed central–satellite configuration.

We find a positive correlation between the flux-weighted velocity dispersion $\tilde{\sigma}$ and stellar mass $M_\star$ (Fig.~\ref{sigma}), with substantial scatter. At fixed $M_\star$, GPs show systematically higher $\tilde{\sigma}$ than their companions, consistent with an excess of kinetic energy in the ionised gas likely driven by the intense stellar feedback associated with their starburst phase. Additionally, estimates of the total dynamical mass for the identified galaxy groups yield values of $\sim 10^{12}$--$10^{13} \, M_\odot$, exceeding the total stellar content by factors of $\sim$250 to $\sim$1300. This large discrepancy could imply the presence of massive dark matter halos and a significant reservoir of gas in the IGM.

Overall, our results are consistent with a picture in which GP starbursts are not primarily triggered by ongoing major mergers with close (10–30 kpc) companions. Instead, a plausible interpretation is that GPs represent transient, extreme episodes of star formation occurring within gas-rich, overdense regions, potentially fuelled by the accretion of gas from a shared large-scale reservoir
\citep[e.g. through cold-mode accretion or filamentary inflows;][]{2005MNRAS.363....2K, 2009MNRAS.395..160K, 2010Natur.467..811C}.
The presence of companions with a wide range of masses and properties, and the observed overdensity of ELGs within $\pm 500$~km~s$^{-1}$, indicate that environment 
likely plays a key role as a provider of fuel and neighbours rather than as a direct major-merger trigger in most cases. However, we caution that our selection is biased towards the star-forming population and
the limited spatial resolution at $z\sim0.2$ (5–15 kpc) implies that very close mergers cannot be ruled out, as systems analogous to the compact GP mergers observed at $z<0.05$ would be only marginally resolved, or even unresolved, in our MUSE data.

\begin{acknowledgements}
Author Antonio Arroyo Polonio acknowledges financial support from the grant CEX2021-001131-S funded by MCIN/AEI/ 10.13039/501100011033 and the ERC synergy grant 101166930 - RECAP. RA acknowledges support from ANID FONDECYT Regular Grant 1202007.
I.B. has received funding from the European Union's Horizon 2020 research and innovation programme under the Marie Sklodowska-Curie Grant agreement ID n.º 101059532. This project was extended for 6 months by the Franziska Seidl Funding Program of the University of Vienna.
JSA is partly funded through grant PID2022-136598NB- C31 (ESTALLIDOS 8) by the Spanish Ministry of Science and Innovation,  “ERDF A way of making Europe”, and  by the European Union through the grant “UNDARK” of the Widening participation and spreading excellence programe (project number 101159929).
MGO acknowledges financial support from the State Agency for Research of the Spanish MCIU through Center of Excellence Severo Ochoa award to the Instituto de Astrofísica de Andalucía CEX2021-001131-S funded by MCIN/AEI/10.13039/501100011033, and from the grant PID2022-136598NB-C32 “Estallidos8”. MGO also acknowledges the support by the project ref. AST22 00001 Subp 11 funded from the EU – NextGenerationEU, PPCC Junta de Andalucía.
\end{acknowledgements}

\bibliographystyle{aa}
\bibliography{main.bib}

@article{cardamone2009galaxy,
  title={\href{https://doi.org/10.1111/j.1365-2966.2009.15383.x}{Galaxy Zoo Green Peas: discovery of a class of compact extremely star-forming galaxies}},
  author={Cardamone, Carolin and Schawinski, Kevin and Sarzi, Marc and Bamford, Steven P and Bennert, Nicola and Urry, C Megan and Lintott, Chris and Keel, William C and Parejko, John and Nichol, Robert C and others},
  journal={\href{https://doi.org/10.1111/j.1365-2966.2009.15383.x}{\mnras}},
  volume={399},
  number={3},
  pages={1191--1205},
  year={2009},
  publisher={The Royal Astronomical Society}
}

@article{vilchez1995spectroscopic,
  title={On the spectroscopic properties of star-forming dwarf galaxies in different environments},
  author={Vilchez, JM},
  journal={\href{https://ui.adsabs.harvard.edu/scan/manifest/1995AJ....110.1090V}{Astronomical Journal}},
  volume={110},
  pages={1090},
  year={1995}
}

@article{perez2011integral,
  title={Integral field spectroscopy of nitrogen overabundant blue compact dwarf galaxies},
  author={P{\'e}rez-Montero, Enrique and V{\'\i}lchez, JM and Cedr{\'e}s, B and H{\"a}gele, Guillermo Federico and Moll{\'a}, M and Kehrig, C and D{\'\i}az, Angeles I and Garc{\'\i}a-Benito, Rub{\'e}n and Mart{\'\i}n-Gord{\'o}n, D},
  journal={\href{https://www.aanda.org/articles/aa/abs/2011/08/aa16582-11/aa16582-11.html}{Astronomy \& Astrophysics}},
  volume={532},
  pages={A141},
  year={2011},
  publisher={EDP Sciences}
}

@article{gehrels1986confidence,
  title={Confidence limits for small numbers of events in astrophysical data},
  author={Gehrels, Neil},
  journal={\href{https://adsabs.harvard.edu/full/1986ApJ...303..336G}{Astrophysical Journal}},
  volume={303},
  pages={336--346},
  year={1986}
}

@ARTICLE{2005MNRAS.363....2K,
       author = {{Kere{\v{s}}}, Du{\v{s}}an and {Katz}, Neal and {Weinberg}, David H. and {Dav{\'e}}, Romeel},
        title = "{How do galaxies get their gas?}",
      journal = {\href{https://academic.oup.com/mnras/article/363/1/2/1300118}{\mnras}},
         year = 2005,
        month = oct,
       volume = {363},
       number = {1},
        pages = {2-28},
          doi = {10.1111/j.1365-2966.2005.09451.x},
archivePrefix = {arXiv},
       eprint = {astro-ph/0407095},
 primaryClass = {astro-ph},
       adsurl = {https://ui.adsabs.harvard.edu/abs/2005MNRAS.363....2K}
}

@ARTICLE{2009MNRAS.395..160K,
       author = {{Kere{\v{s}}}, Du{\v{s}}an and {Katz}, Neal and {Fardal}, Mark and {Dav{\'e}}, Romeel and {Weinberg}, David H.},
        title = "{Galaxies in a simulated {\ensuremath{\Lambda}}CDM Universe - I. Cold mode and hot cores}",
      journal = {\href{https://academic.oup.com/mnras/article/395/1/160/1079146}{\mnras}},
         year = 2009,
        month = may,
       volume = {395},
       number = {1},
        pages = {160-179},
          doi = {10.1111/j.1365-2966.2009.14541.x},
archivePrefix = {arXiv},
       eprint = {0809.1430},
 primaryClass = {astro-ph},
       adsurl = {https://ui.adsabs.harvard.edu/abs/2009MNRAS.395..160K}
}

@ARTICLE{2010Natur.467..811C,
       author = {{Cresci}, G. and {Mannucci}, F. and {Maiolino}, R. and {Marconi}, A. and {Gnerucci}, A. and {Magrini}, L.},
        title = "{Gas accretion as the origin of chemical abundance gradients in distant galaxies}",
      journal = {\href{https://www.nature.com/articles/nature09451}{\nat}},
         year = 2010,
        month = oct,
       volume = {467},
       number = {7317},
        pages = {811-813},
          doi = {10.1038/nature09451},
archivePrefix = {arXiv},
       eprint = {1010.2534},
 primaryClass = {astro-ph.CO},
       adsurl = {https://ui.adsabs.harvard.edu/abs/2010Natur.467..811C}
}

@article{bruzual2003stellar,
  title={Stellar population synthesis at the resolution of 2003},
  author={Bruzual, Gustavo and Charlot, Stephane},
  journal={\href{https://academic.oup.com/mnras/article/344/4/1000/968846?guestAccessKey=}{Monthly Notices of the Royal Astronomical Society}},
  volume={344},
  number={4},
  pages={1000--1028},
  year={2003},
  publisher={Blackwell Science Ltd Oxford, UK}
}

@article{noeske2001faint,
  title={On faint companions in the close environment of star-forming dwarf galaxies-Possible external star formation triggers?},
  author={Noeske, KG and Iglesias-P{\'a}ramo, J and V{\'\i}lchez, JM and Papaderos, P and Fricke, KJ},
  journal={\href{https://www.aanda.org/articles/aa/full/2001/21/aa10303/node5.html}{Astronomy \& Astrophysics}},
  volume={371},
  number={3},
  pages={806--815},
  year={2001},
  publisher={EDP Sciences}
}

@article{pustilnik2001environment,
  title={Environment status of blue compact galaxies and trigger of star formation},
  author={Pustilnik, Simon A and Kniazev, Alexei Y and Lipovetsky, Valentin A and Ugryumov, AV},
  journal={\href{https://www.aanda.org/articles/aa/full/2001/25/aa10617/node3.html}{Astronomy \& Astrophysics}},
  volume={373},
  number={1},
  pages={24--37},
  year={2001},
  publisher={EDP Sciences}
}

@article{telles2000local,
  title={The local environment of H ii galaxies},
  author={Telles, Eduardo and Maddox, Steve},
  journal={\href{https://academic.oup.com/mnras/article/311/2/307/964327?guestAccessKey=}{Monthly Notices of the Royal Astronomical Society}},
  volume={311},
  number={2},
  pages={307--312},
  year={2000},
  publisher={Blackwell Science Ltd Oxford, UK}
}

@article{campos1993empirical,
  title={Empirical characterization of blue dwarf galaxies},
  author={Campos-Aguilar, Ana and Moles, Mariano and Masegosa, Josefa},
  journal={\href{https://ui.adsabs.harvard.edu/scan/manifest/1993AJ....106.1784C}{Astronomical Journal}},
  volume={106},
  pages={1784--1796},
  year={1993}
}

@article{gupta2026blueberry,
  title={Blueberry and Green Pea galaxies live in low density environments},
  author={Gupta, Maitrayee and Svoboda, Ji{\v{r}}{\'\i} and Kouroumpatzakis, Konstantinos and Peschken, Nicolas and Boorman, Peter G and Borkar, Abhijeet},
  journal={\href{https://arxiv.org/abs/2604.26066}{arXiv}},
  year={2026}
}

@article{chabrier2003galactic,
  title={Galactic stellar and substellar initial mass function},
  author={Chabrier, Gilles},
  journal={\href{https://iopscience.iop.org/article/10.1086/376392/meta?casa_token=PECKWHg4tEAAAAAA:t0-2NEGXLEplk5neBj-xb27MzlU3pojPNTexb5m3uIxjQ_53Qebo32xWjeRxctd6RDl5I_fckBPK9EMlGIpnQVLXxQ&casa_token=1ujfthEz3wQAAAAA:_M1zjttlgs40BbsJobmHW365lAoQ_4UC0QXt1i51RW8Ee4b1itohe1YRR0VT6O2_tLe-mg13kQcmimYKLqQG6-vksA}{Publications of the Astronomical Society of the Pacific}},
  volume={115},
  number={809},
  pages={763--795},
  year={2003},
  publisher={The University of Chicago Press}
}

@article{calzetti2000dust,
  title={The dust content and opacity of actively star-forming galaxies},
  author={Calzetti, Daniela and Armus, Lee and Bohlin, Ralph C and Kinney, Anne L and Koornneef, Jan and Storchi-Bergmann, Thaisa},
  journal={\href{https://iopscience.iop.org/article/10.1086/308692/meta}{The Astrophysical Journal}},
  volume={533},
  number={2},
  pages={682--695},
  year={2000}
}

@article{kouroumpatzakis2024blueberry,
  title={Blueberry galaxies up to 200 Mpc and their optical and infrared properties},
  author={Kouroumpatzakis, K and Svoboda, J and Zezas, A and Borkar, A and Kyritsis, E and Boorman, PG and Daoutis, C and Adamcov{\'a}, B and Grossov{\'a}, R},
  journal={\href{https://www.aanda.org/articles/aa/abs/2024/08/aa49766-24/aa49766-24.html}{Astronomy \& Astrophysics}},
  volume={688},
  pages={A159},
  year={2024},
  publisher={EDP Sciences}
}

@article{almeida2013local,
  title={Local tadpole galaxies: dynamics and metallicity},
  author={Sánchez Almeida, J. and Mu{\~n}oz-Tu{\~n}{\'o}n, C and Elmegreen, Debra Meloy and Elmegreen, Bruce G and M{\'e}ndez-Abreu, J},
  journal={\href{https://iopscience.iop.org/article/10.1088/0004-637X/767/1/74/meta}{The Astrophysical Journal}},
  volume={767},
  number={1},
  pages={74},
  year={2013},
  publisher={IOP Publishing}
}

@article{lumbreras2022j,
  title={J-PLUS: Uncovering a large population of extreme [OIII] emitters in the local Universe},
  author={Lumbreras-Calle, A and L{\'o}pez-Sanjuan, Carlos and Sobral, D and Fern{\'a}ndez-Ontiveros, JA and Vilchez, JM and Hern{\'a}n-Caballero, Antonio and Akhlaghi, M and Diaz-Garcia, LA and Alcaniz, Jailson and Angulo, RE and others},
  journal={\href{https://www.aanda.org/articles/aa/full_html/2022/12/aa42898-21/aa42898-21.html}{Astronomy \& Astrophysics}},
  volume={668},
  pages={A60},
  year={2022},
  publisher={EDP Sciences}
}

@article{amorin2012star,
  title={The star formation history and metal content of the green peas. New detailed gtc-osiris spectrophotometry of three galaxies},
  author={Amor{\'\i}n, Ricardo and P{\'e}rez-Montero, E and V{\'\i}lchez, JM and Papaderos, P},
  journal={\href{https://iopscience.iop.org/article/10.1088/0004-637X/749/2/185/meta}{The Astrophysical Journal}},
  volume={749},
  number={2},
  pages={185},
  year={2012},
  publisher={IOP Publishing}
}

@inproceedings{bacon2010muse,
  title={The MUSE second-generation VLT instrument},
  author={Bacon, R and Accardo, M and Adjali, L and Anwand, H and Bauer, S and Biswas, I and Blaizot, J and Boudon, D and Brau-Nogue, S and Brinchmann, J and others},
  booktitle={\href{https://www.spiedigitallibrary.org/conference-proceedings-of-spie/7735/773508/The-MUSE-second-generation-VLT-instrument/10.1117/12.856027.short?SSO=1}{Ground-based and Airborne Instrumentation for Astronomy III}},
  volume={7735},
  pages={131--139},
  year={2010},
  organization={SPIE}
}

@article{cardelli1989relationship,
  title={The relationship between infrared, optical, and ultraviolet extinction},
  author={Cardelli, Jason A and Clayton, Geoffrey C and Mathis, John S},
  journal={\href{https://adsabs.harvard.edu/pdf/1989ApJ...345..245C7}{The Astrophysical Journal}},
  volume={345},
  pages={245--256},
  year={1989}
}

@article{lofthouse2017local,
  title={Local analogues of high-redshift star-forming galaxies: integral field spectroscopy of green peas},
  author={Lofthouse, Emma K and Houghton, Ryan CW and Kaviraj, Sugata},
  journal={\href{https://academic.oup.com/mnras/article/471/2/2311/3930848?login=true}{\mnras}},
  volume={471},
  number={2},
  pages={2311--2320},
  year={2017},
  publisher={Oxford University Press}
}

@article{kennicutt1998global,
  title={The global Schmidt law in star-forming galaxies},
  author={Kennicutt Jr, Robert C},
  journal={\href{https://iopscience.iop.org/article/10.1086/305588/meta}{The astrophysical journal}},
  volume={498},
  number={2},
  pages={541},
  year={1998},
  publisher={IOP Publishing}
}

@article{weilbacher2020data,
  title={The data processing pipeline for the MUSE instrument},
  author={Weilbacher, Peter M and Palsa, Ralf and Streicher, Ole and Bacon, Roland and Urrutia, Tanya and Wisotzki, Lutz and Conseil, Simon and Husemann, Bernd and Jarno, Aur{\'e}lien and Kelz, Andreas and others},
  journal={\href{https://www.aanda.org/articles/aa/abs/2020/09/aa37855-20/aa37855-20.html}{\aap}},
  volume={641},
  pages={A28},
  year={2020},
  publisher={EDP Sciences}
}

@article{whitaker2012star,
  title={The star formation mass sequence out to z= 2.5},
  author={Whitaker, Katherine E and Van Dokkum, Pieter G and Brammer, Gabriel and Franx, Marijn},
  journal={\href{https://iopscience.iop.org/article/10.1088/2041-8205/754/2/L29/meta}{The Astrophysical Journal Letters}},
  volume={754},
  number={2},
  pages={L29},
  year={2012},
  publisher={IOP Publishing}
}

@article{amorin2010oxygen,
  title={On the oxygen and nitrogen chemical abundances and the evolution of the “green pea” galaxies},
  author={Amor{\'\i}n, Ricardo O and P{\'e}rez-Montero, Enrique and V{\'\i}lchez, JM},
  journal={\href{https://iopscience.iop.org/article/10.1088/2041-8205/715/2/L128/meta}{The Astrophysical Journal Letters}},
  volume={715},
  number={2},
  pages={L128},
  year={2010},
  publisher={IOP Publishing}
}

@ARTICLE{perez2014code,
       author = {{P{\'e}rez-Montero}, E.},
        title = "{Deriving model-based T$_{e}$-consistent chemical abundances in ionized gaseous nebulae}",
      journal = {\href{https://academic.oup.com/mnras/article/441/3/2663/1133246}{\mnras}},
         year = 2014,
        month = jul,
       volume = {441},
       number = {3},
        pages = {2663-2675},
          doi = {10.1093/mnras/stu753},
archivePrefix = {arXiv},
       eprint = {1404.3936},
 primaryClass = {astro-ph.GA},
       adsurl = {https://ui.adsabs.harvard.edu/abs/2014MNRAS.441.2663P}
}

@article{yang2017lyalpha,
  title={Ly$\alpha$ profile, dust, and prediction of Ly$\alpha$ escape fraction in green pea galaxies},
  author={Yang, Huan and Malhotra, Sangeeta and Gronke, Max and Rhoads, James E and Leitherer, Claus and Wofford, Aida and Jiang, Tianxing and Dijkstra, Mark and Tilvi, V and Wang, Junxian},
  journal={\href{https://iopscience.iop.org/article/10.3847/1538-4357/aa7d4d/meta}{The Astrophysical Journal}},
  volume={844},
  number={2},
  pages={171},
  year={2017},
  publisher={IOP Publishing}
}

@article{paudel2024discovery,
  title={Discovery of a Rare Group of Dwarf Galaxies in the Local Universe},
  author={Paudel, Sanjaya and Sabiu, Cristiano G and Yoon, Suk-Jin and Duc, Pierre-Alain and Yoo, Jaewon and M{\"u}ller, Oliver},
  journal={\href{https://iopscience.iop.org/article/10.3847/2041-8213/ad8f3c/meta}{The Astrophysical Journal Letters}},
  volume={976},
  number={1},
  pages={L18},
  year={2024},
  publisher={IOP Publishing}
}

@article{heisler1985estimating,
  title={Estimating the masses of galaxy groups-Alternatives to the virial theorem},
  author={Heisler, Julia and Tremaine, Scott and Bahcall, John N},
  journal={\href{https://adsabs.harvard.edu/full/record/seri/ApJ../0298/1985ApJ...298....8H.html}{Astrophysical Journal, Part 1 (ISSN 0004-637X), vol. 298, Nov. 1, 1985, p. 8-17.}},
  volume={298},
  pages={8--17},
  year={1985}
}

@article{ashley2014h,
  title={THE H i CHRONICLES OF LITTLE THINGS BCDs II: THE ORIGIN OF IC 10's H i STRUCTURE},
  author={Ashley, Trisha and Elmegreen, Bruce G and Johnson, Megan and Nidever, David L and Simpson, Caroline E and Pokhrel, Nau Raj},
  journal={\href{https://iopscience.iop.org/article/10.1088/0004-6256/148/6/130/meta}{The Astronomical Journal}},
  volume={148},
  number={6},
  pages={130},
  year={2014},
  publisher={IOP Publishing}
}

@article{almeida2015localized,
  title={Localized starbursts in dwarf galaxies produced by the impact of low-metallicity cosmic gas clouds},
  author={Sánchez Almeida, J. and Elmegreen, Bruce G and Mu{\~n}oz-Tu{\~n}{\'o}n, C and Elmegreen, Debra Meloy and P{\'e}rez-Montero, E and Amor{\'\i}n, R and Ascasibar, Y and Papaderos, P and V{\'\i}lchez, JM and others},
  journal={\href{https://iopscience.iop.org/article/10.1088/2041-8205/810/2/L15/meta}{The Astrophysical Journal Letters}},
  volume={810},
  number={2},
  pages={L15},
  year={2015},
  publisher={IOP Publishing}
}

@article{sanchez2014star,
  title={Star formation sustained by gas accretion},
  author={S{\'a}nchez Almeida, Jorge and Elmegreen, Bruce G and Munoz-Tun{\'o}n, Casiana and Elmegreen, Debra Meloy},
  journal={\href{https://link.springer.com/article/10.1007/s00159-014-0071-1}{The Astronomy and Astrophysics Review}},
  volume={22},
  number={1},
  pages={71},
  year={2014},
  publisher={Springer}
}

@article{dekel2006galaxy,
  title={Galaxy bimodality due to cold flows and shock heating},
  author={Dekel, Avishai and Birnboim, Yuval},
  journal={\href{https://academic.oup.com/mnras/article/368/1/2/968063}{Monthly notices of the royal astronomical society}},
  volume={368},
  number={1},
  pages={2--20},
  year={2006},
  publisher={Blackwell Publishing Ltd Oxford, UK}
}

@article{dekel2009cold,
  title={Cold streams in early massive hot haloes as the main mode of galaxy formation},
  author={Dekel, A and Birnboim, Y and Engel, G and Freundlich, J and Goerdt, T and Mumcuoglu, M and Neistein, E and Pichon, C and Teyssier, R and Zinger, E},
  journal={\href{https://www.nature.com/articles/nature07648}{Nature}},
  volume={457},
  number={7228},
  pages={451--454},
  year={2009},
  publisher={Nature Publishing Group UK London}
}

@article{zitrin2009star,
  title={Star formation properties of isolated blue compact galaxies},
  author={Zitrin, Adi and Brosch, Noah and Bilenko, Benny},
  journal={\href{https://academic.oup.com/mnras/article/399/2/924/1063197}{Monthly Notices of the Royal Astronomical Society}},
  volume={399},
  number={2},
  pages={924--933},
  year={2009},
  publisher={Blackwell Publishing Ltd Oxford, UK}
}

@article{purkayastha2022green,
  title={A Green Pea Starburst Arising from a Galaxy--Galaxy Merger},
  author={Purkayastha, S and Kanekar, N and Chengalur, JN and Malhotra, S and Rhoads, J and Ghosh, T},
  journal={\href{https://iopscience.iop.org/article/10.3847/2041-8213/ac7522/meta}{The Astrophysical Journal Letters}},
  volume={933},
  number={1},
  pages={L11},
  year={2022},
  publisher={IOP Publishing}
}

@article{rawat2007unravelling,
  title={Unravelling the morphologies of luminous compact galaxies using the HST/ACS GOODS survey},
  author={Rawat, A and Kembhavi, Ajit K and Hammer, F and Flores, H and Barway, S},
  journal={\href{https://www.aanda.org/articles/aa/abs/2007/26/aa5737-06/aa5737-06.html}{Astronomy \& Astrophysics}},
  volume={469},
  number={2},
  pages={483--501},
  year={2007},
  publisher={EDP Sciences}
}

@article{purkayastha2024second,
  title={The Second Case of a Major Merger Triggering a Starburst in a Green Pea Galaxy},
  author={Purkayastha, S and Kanekar, N and Kumari, S and Rhoads, J and Malhotra, S and Pharo, J and Ghosh, T},
  journal={\href{https://iopscience.iop.org/article/10.3847/1538-4357/ad8dd2/meta}{The Astrophysical Journal}},
  volume={977},
  number={1},
  pages={68},
  year={2024},
  publisher={IOP Publishing}
}

@article{izotov2016eight,
  title={Eight per cent leakage of Lyman continuum photons from a compact, star-forming dwarf galaxy},
  author={Izotov, YI and Orlitov{\'a}, I and Schaerer, D and Thuan, TX and Verhamme, A and Guseva, NG and Worseck, G},
  journal={\href{https://www.nature.com/articles/nature16456}{Nature}},
  volume={529},
  number={7585},
  pages={178--180},
  year={2016},
  publisher={Nature Publishing Group UK London}
}

@article{baldwin1981classification,
  title={Classification parameters for the emission-line spectra of extragalactic objects.},
  author={Baldwin, Jack A and Phillips, Mark M and Terlevich, Roberto},
  journal={\href{https://iopscience.iop.org/article/10.1086/130766/meta?casa_token=evWyI8PdVpEAAAAA:Oc2gmARQyGXnYxCYCyVE4ptSSnylIxe6ruVZcM2Hzod3ZmdrReVP0SQrJoqrcHfXLABRze7OgYgF}{Publications of the Astronomical Society of the Pacific}},
  volume={93},
  number={551},
  pages={5},
  year={1981},
  publisher={IOP Publishing}
}

@article{kewley2006host,
  title={The host galaxies and classification of active galactic nuclei},
  author={Kewley, Lisa J and Groves, Brent and Kauffmann, Guinevere and Heckman, Tim},
  journal={\href{https://academic.oup.com/mnras/article/372/3/961/972572?login=true}{\mnras}},
  volume={372},
  number={3},
  pages={961--976},
  year={2006},
  publisher={Blackwell Publishing Ltd Oxford, UK}
}

@article{gimenez2025j,
  title={J-PAS: First Identification, Physical Properties and Ionization Efficiency of Extreme Emission Line Galaxies},
  author={Gim{\'e}nez-Alc{\'a}zar, A and Amor{\'\i}n, R and V{\'\i}lchez, JM and Hern{\'a}n-Caballero, A and Gonz{\'a}lez-Otero, M and Arroyo-Polonio, A and Iglesias-P{\'a}ramo, J and Lumbreras-Calle, A and Fern{\'a}ndez-Ontiveros, JA and Bonatto, L and others},
  journal={\href{https://arxiv.org/abs/2512.08484}{arXiv preprint arXiv:2512.08484}},
  year={2025}
}

@article{laufman2022triggering,
  title={On the triggering of extreme starburst events in low-metallicity galaxies: a deep search for companions of Green Peas},
  author={Laufman, Lauren and Scarlata, Claudia and Hayes, Matthew and Skillman, Evan},
  journal={\href{https://iopscience.iop.org/article/10.3847/1538-4357/ac97ef/meta}{The Astrophysical Journal}},
  volume={940},
  number={1},
  pages={31},
  year={2022},
  publisher={IOP Publishing}
}

@article{billand2025investigating,
  title={Investigating the Growth of Little Red Dot Descendants at z< 4 with the JWST},
  author={Billand, Jean-Baptiste and Elbaz, David and Gentile, Fabrizio and Tarrasse, Maxime and Franco, Maximilien and Magnelli, Benjamin and Daddi, Emanuele and Lyu, Yipeng and Dekel, Avishai and Pacucci, Fabio and others},
  journal={\href{https://www.aanda.org/component/article?access=doi&doi=10.1051/0004-6361/202556303}{Astronomy \& Astrophysics}},
  year={2025},
  publisher={EDP Sciences}
}

@article{puertas2022mass,
  title={Mass--metallicity and star formation rate in galaxies: A complex relation tuned to stellar age},
  author={Puertas, S Duarte and Vilchez, JM and Iglesias-P{\'a}ramo, J and Moll{\'a}, M and P{\'e}rez-Montero, Enrique and Kehrig, C and Pilyugin, LS and Zinchenko, IA},
  journal={\href{https://www.aanda.org/articles/aa/abs/2022/10/aa41571-21/aa41571-21.html}{Astronomy \& Astrophysics}},
  volume={666},
  pages={A186},
  year={2022},
  publisher={EDP Sciences}
}

@ARTICLE{lineaBPT1,
       author = {{Kewley}, L.~J. and {Dopita}, M.~A. and {Sutherland}, R.~S. and {Heisler}, C.~A. and {Trevena}, J.},
        title = "{Theoretical Modeling of Starburst Galaxies}",
      journal = {\href{https://iopscience.iop.org/article/10.1086/321545}{\apj}},
         year = 2001,
        month = jul,
       volume = {556},
       number = {1},
        pages = {121-140},
          doi = {10.1086/321545},
archivePrefix = {arXiv},
       eprint = {astro-ph/0106324},
 primaryClass = {astro-ph},
       adsurl = {https://ui.adsabs.harvard.edu/abs/2001ApJ...556..121K}
}

@ARTICLE{lineaBPT2,
       author = {{Kauffmann}, Guinevere and {Heckman}, Timothy M. and {Tremonti}, Christy and {Brinchmann}, Jarle and {Charlot}, St{\'e}phane and {White}, Simon D.~M. and {Ridgway}, Susan E. and {Brinkmann}, Jon and {Fukugita}, Masataka and {Hall}, Patrick B. and {Ivezi{\'c}}, {\v{Z}}eljko and {Richards}, Gordon T. and {Schneider}, Donald P.},
        title = "{The host galaxies of active galactic nuclei}",
      journal = {\href{https://academic.oup.com/mnras/article/346/4/1055/1062435}{\mnras}},
         year = 2003,
        month = dec,
       volume = {346},
       number = {4},
        pages = {1055-1077},
          doi = {10.1111/j.1365-2966.2003.07154.x},
archivePrefix = {arXiv},
       eprint = {astro-ph/0304239},
 primaryClass = {astro-ph},
       adsurl = {https://ui.adsabs.harvard.edu/abs/2003MNRAS.346.1055K}
}

@article{efron1985bootstrap,
  title={The bootstrap method for assessing statistical accuracy},
  author={Efron, Bradley and Tibshirani, Robert},
  journal={\href{https://link.springer.com/article/10.2333/bhmk.12.17_1}{Behaviormetrika}},
  volume={12},
  pages={1--35},
  year={1985},
  publisher={Springer}
}

@article{perez2021extreme,
  title={Extreme emission-line galaxies in SDSS--I. Empirical and model-based calibrations of chemical abundances},
  author={P{\'e}rez-Montero, Enrique and Amor{\'\i}n, R and S{\'a}nchez Almeida, J and V{\'\i}lchez, JM and Garc{\'\i}a-Benito, Rub{\'e}n and Kehrig, C},
  journal={\href{https://academic.oup.com/mnras/article-abstract/504/1/1237/6195523}{\mnras}},
  volume={504},
  number={1},
  pages={1237--1252},
  year={2021},
  publisher={Oxford University Press}
}

@article{breda2022characterisation,
  title={Characterisation of the stellar content of SDSS EELGs through self-consistent spectral modelling},
  author={Breda, Iris and Vilchez, Jos{\'e} M and Papaderos, Polychronis and Cardoso, Leandro and Amorin, Ricardo O and Arroyo-Polonio, Antonio and Iglesias-P{\'a}ramo, Jorge and Kehrig, Carolina and P{\'e}rez-Montero, Enrique},
  journal={\href{https://www.aanda.org/articles/aa/abs/2022/07/aa42805-21/aa42805-21.html}{\aap}},
  volume={663},
  pages={A29},
  year={2022},
  publisher={EDP Sciences}
}

@article{iglesias2022minijpas,
  title={The miniJPAS survey: A search for extreme emission-line galaxies},
  author={Iglesias-P{\'a}ramo, J and Arroyo, A and Kehrig, C and V{\'\i}lchez, JM and Puertas, S Duarte and P{\'e}rez-Montero, Enrique and Breda, I and Jim{\'e}nez-Teja, Y and Sanjuan, C Lopez and Lumbreras-Calle, A and others},
  journal={\href{https://www.aanda.org/articles/aa/abs/2022/09/aa43931-22/aa43931-22.html}{\aap}},
  volume={665},
  pages={A95},
  year={2022},
  publisher={EDP Sciences}
}

@article{molla2009popstar,
  title={PopStar I: evolutionary synthesis model description},
  author={Moll{\'a}, M and Garc{\'\i}a-Vargas, ML and Bressan, A},
  journal={\href{https://academic.oup.com/mnras/article/398/1/451/1099698}{Monthly Notices of the Royal Astronomical Society}},
  volume={398},
  number={1},
  pages={451--470},
  year={2009},
  publisher={Blackwell Publishing Ltd Oxford, UK}
}

@article{arroyo2023muse,
  title={A MUSE/VLT spatially resolved study of the emission structure of Green Pea galaxies},
  author={Arroyo-Polonio, A and Iglesias-P{\'a}ramo, J and Kehrig, C and V{\'\i}lchez, JM and Amor{\'\i}n, R and Breda, I and P{\'e}rez-Montero, Enrique and P{\'e}rez-D{\'\i}az, B and Hayes, M},
  journal={\href{https://www.aanda.org/articles/aa/abs/2023/09/aa46192-23/aa46192-23.html}{Astronomy \& Astrophysics}},
  volume={677},
  pages={A114},
  year={2023},
  publisher={EDP Sciences}
}

@article{perez2024departure,
  title={A departure from the mass--metallicity relation in merging galaxies due to an infall of metal-poor gas},
  author={P{\'e}rez-D{\'\i}az, Borja and P{\'e}rez-Montero, Enrique and Fern{\'a}ndez-Ontiveros, Juan A and V{\'\i}lchez, Jos{\'e} M and Amor{\'\i}n, Ricardo},
  journal={\href{https://www.nature.com/articles/s41550-023-02171-x}{Nature Astronomy}},
  volume={8},
  number={3},
  pages={368--376},
  year={2024},
  publisher={Nature Publishing Group UK London}
}

@article{veilleux2002optical,
  title={Optical and near-infrared imaging of the IRAS 1 Jy sample of ultraluminous infrared galaxies. II. The analysis},
  author={Veilleux, S and Kim, D-C and Sanders, DB},
  journal={\href{https://iopscience.iop.org/article/10.1086/343844/meta}{The Astrophysical Journal Supplement Series}},
  volume={143},
  number={2},
  pages={315},
  year={2002},
  publisher={IOP Publishing}
}

@article{sanders1996luminous,
  title={Luminous infrared galaxies},
  author={Sanders, DB and Mirabel, IF},
  journal={\href{https://www.annualreviews.org/content/journals/10.1146/annurev.astro.34.1.749}{Annual Review of Astronomy and Astrophysics}},
  volume={34},
  number={1},
  pages={749--792},
  year={1996},
  publisher={Annual Reviews 4139 El Camino Way, PO Box 10139, Palo Alto, CA 94303-0139, USA}
}

@article{ellison2008galaxy,
  title={Galaxy pairs in the sloan digital sky survey. i. star formation, active galactic nucleus fraction, and the luminosity/mass--metallicity relation},
  author={Ellison, Sara L and Patton, David R and Simard, Luc and McConnachie, Alan W},
  journal={\href{https://iopscience.iop.org/article/10.1088/0004-6256/135/5/1877/meta}{The Astronomical Journal}},
  volume={135},
  number={5},
  pages={1877},
  year={2008},
  publisher={IOP Publishing}
}

@article{barton2000tidally,
  title={Tidally Triggered Star Formation in Close Pairsof Galaxies},
  author={Barton, Elizabeth J and Geller, Margaret J and Kenyon, Scott J},
  journal={\href{https://iopscience.iop.org/article/10.1086/308392/meta}{The Astrophysical Journal}},
  volume={530},
  number={2},
  pages={660},
  year={2000},
  publisher={IOP Publishing}
}

@article{kennicutt1987effects,
  title={The effects of interactions on spiral galaxies. II-Disk star-formation rates},
  author={Kennicutt Jr, Robert C and Keel, William C and Van der Hulst, JM and Hummel, E and Roettiger, Kurt A},
  journal={\href{https://adsabs.harvard.edu/full/record/seri/AJ.../0093/1987AJ.....93.1011K.html}{Astronomical Journal (ISSN 0004-6256), vol. 93, May 1987, p. 1011-1023.}},
  volume={93},
  pages={1011--1023},
  year={1987}
}

@article{boquien2019cigale,
  title={CIGALE: a Python code investigating galaxy emission},
  author={Boquien, M and Burgarella, D and Roehlly, Y and Buat, V and Ciesla, L and Corre, D and Inoue, AK and Salas, H},
  journal={\href{https://www.aanda.org/articles/aa/abs/2019/02/aa34156-18/aa34156-18.html}{Astronomy \& Astrophysics}},
  volume={622},
  pages={A103},
  year={2019},
  publisher={EDP Sciences}
}

@article{kennicutt2012star,
  title={Star formation in the Milky Way and nearby galaxies},
  author={Kennicutt Jr, Robert C and Evans, Neal J},
  journal={\href{https://www.annualreviews.org/content/journals/10.1146/annurev-astro-081811-125610}{Annual Review of Astronomy and Astrophysics}},
  volume={50},
  number={1},
  pages={531--608},
  year={2012},
  publisher={Annual Reviews}
}

@article{joseph1985recent,
  title={Recent star formation in interacting galaxies--II. Super starbursts in merging galaxies},
  author={Joseph, RD and Wright, GS},
  journal={\href{https://academic.oup.com/mnras/article/214/2/87/1031920}{Monthly Notices of the Royal Astronomical Society}},
  volume={214},
  number={2},
  pages={87--95},
  year={1985},
  publisher={Oxford University Press Oxford, UK}
}

@article{izotov2016detection,
  title={Detection of high Lyman continuum leakage from four low-redshift compact star-forming galaxies},
  author={Izotov, YI and Schaerer, D and Thuan, TX and Worseck, G and Guseva, NG and Orlitov{\'a}, I and Verhamme, A},
  journal={\href{https://academic.oup.com/mnras/article/461/4/3683/2608537?login=false}{\mnras}},
  volume={461},
  number={4},
  pages={3683--3701},
  year={2016},
  publisher={Oxford University Press}
}

@article{izotov2011green,
  title={Green pea galaxies and cohorts: luminous compact emission-line galaxies in the sloan digital sky survey},
  author={Izotov, Yuri I and Guseva, Natalia G and Thuan, Trinh X},
  journal={\href{https://iopscience.iop.org/article/10.1088/0004-637X/728/2/161/meta}{The Astrophysical Journal}},
  volume={728},
  number={2},
  pages={161},
  year={2011},
  publisher={IOP Publishing}
}

\begin{appendix}

\section{FoV of the GPs with companions}
\label{appendix}

\begin{figure*}[h!]
\centering
   \includegraphics[width=\textwidth]{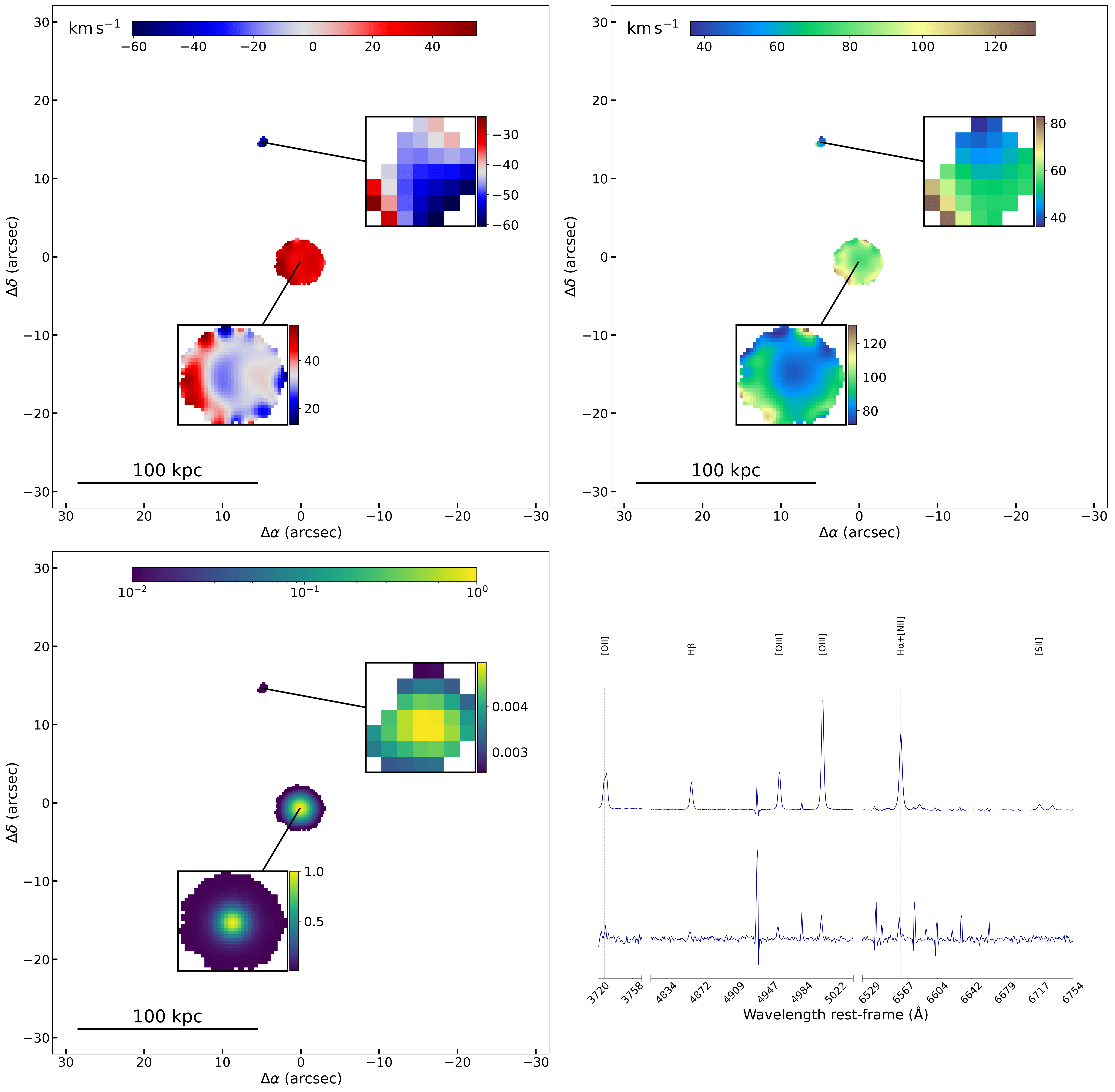}
     \caption{GP3 system. For details of the meaning of each panel, see Fig. \ref{panel_GP1}}
     \label{panel_GP3}
\end{figure*}

\begin{figure*}[h!]
\centering
   \includegraphics[width=\textwidth]{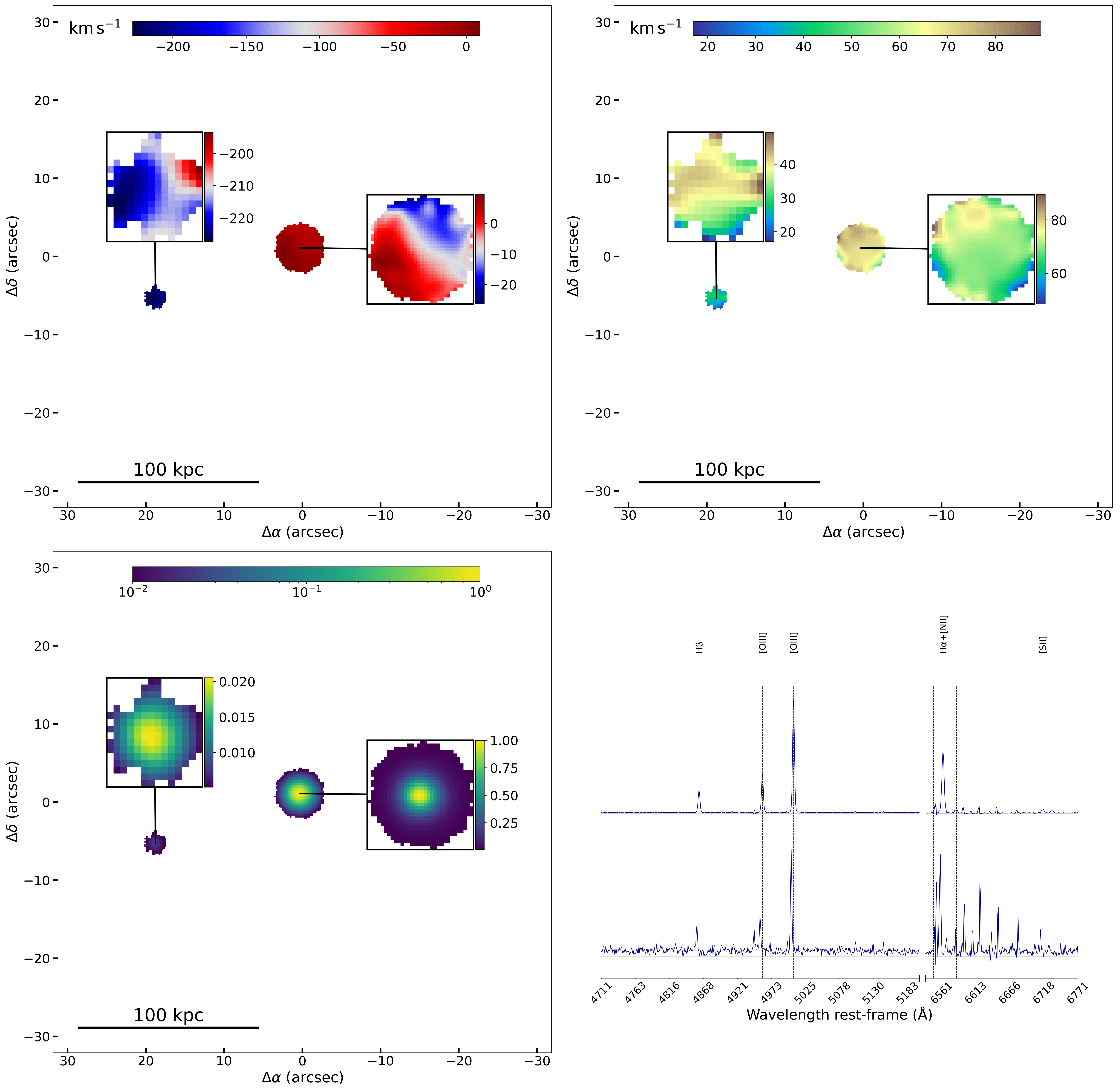}
     \caption{GP14 system. For details of the meaning of each panel, see Fig. \ref{panel_GP1}}
     \label{panel_GP14}
\end{figure*}

\begin{figure*}[h!]
\centering
   \includegraphics[width=\textwidth]{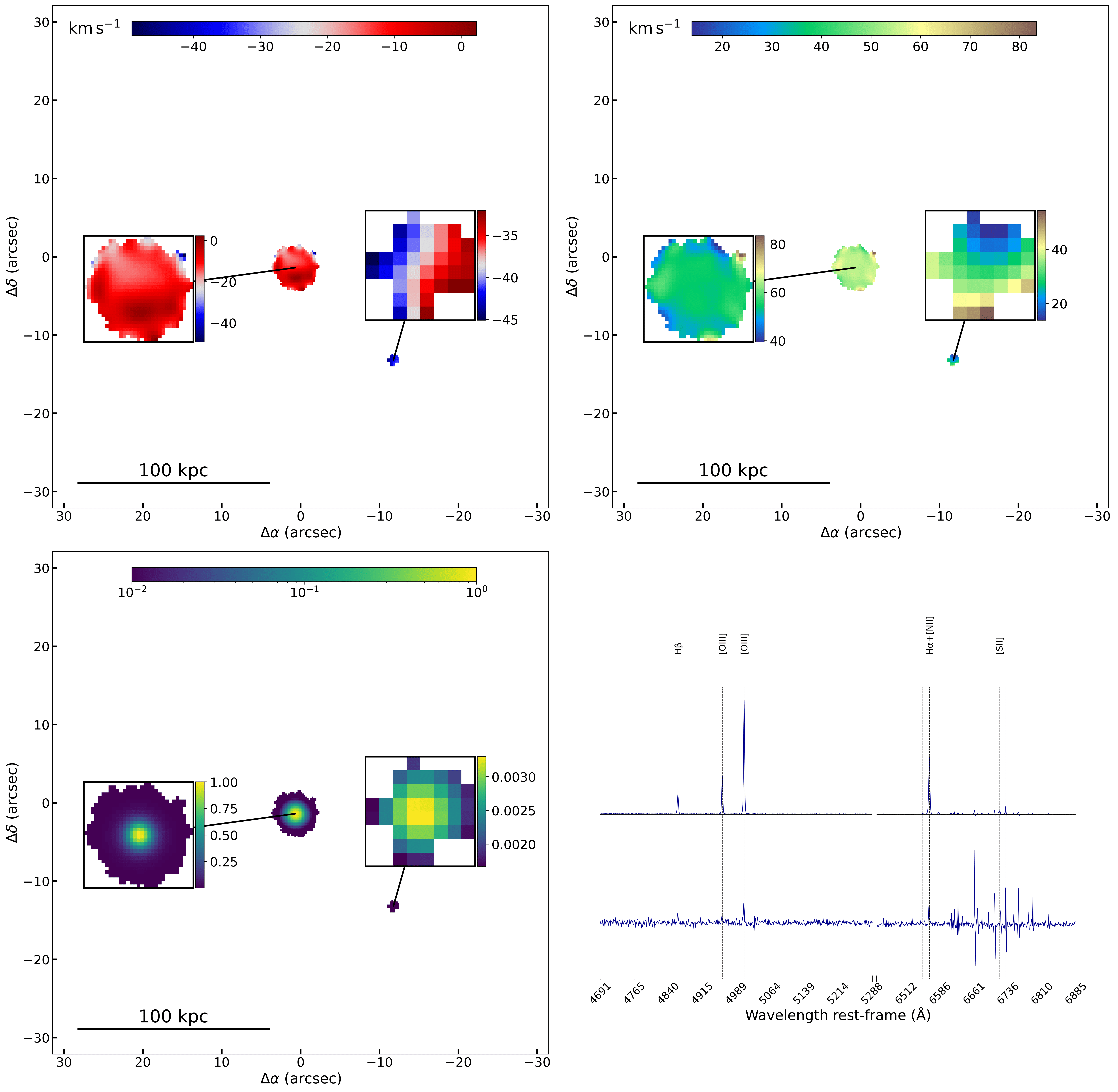}
     \caption{GP16 system. For details of the meaning of each panel, see Fig. \ref{panel_GP1}}
     \label{panel_GP16}
\end{figure*}

\begin{figure*}[h!]
\centering
   \includegraphics[width=\textwidth]{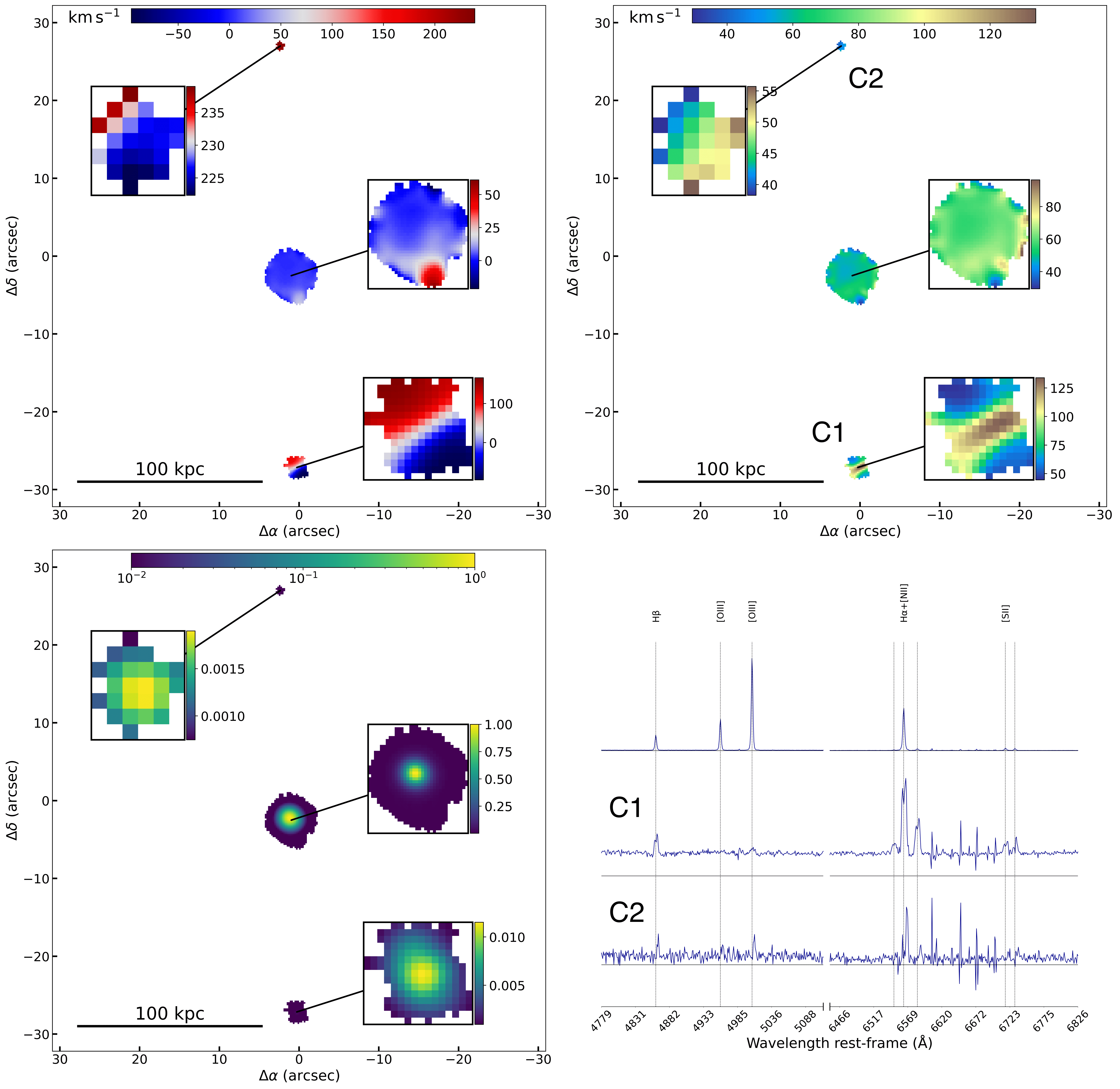}
     \caption{GP17 system. For details of the meaning of each panel, see Fig. \ref{panel_GP1}}
     \label{panel_GP17}
\end{figure*}

\begin{figure*}[h!]
\centering
   \includegraphics[width=\textwidth]{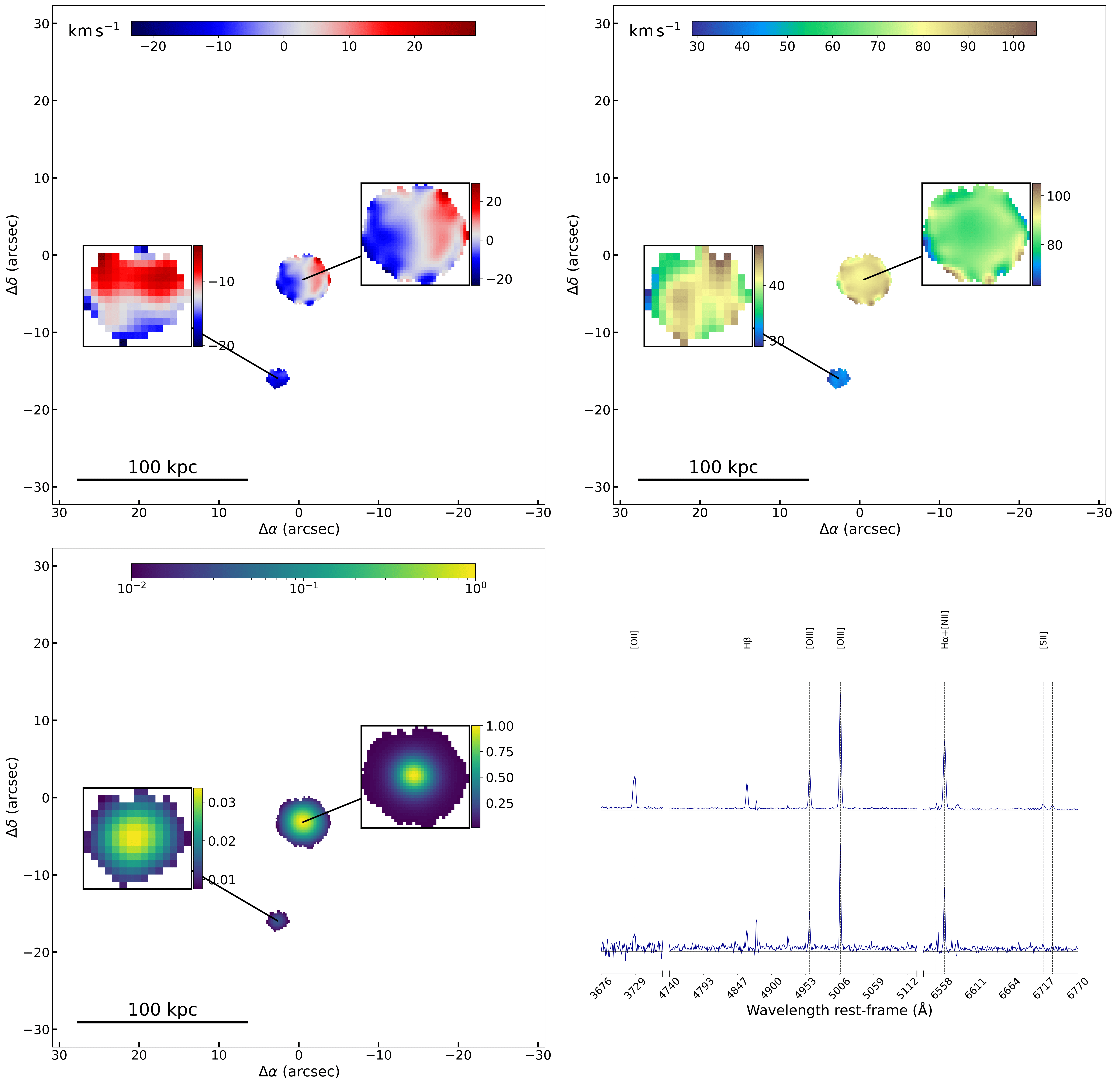}
     \caption{GP21 system. For details of the meaning of each panel, see Fig. \ref{panel_GP1}}
     \label{panel_GP21}
\end{figure*}

\begin{figure*}[h!]
\centering
   \includegraphics[width=\textwidth]{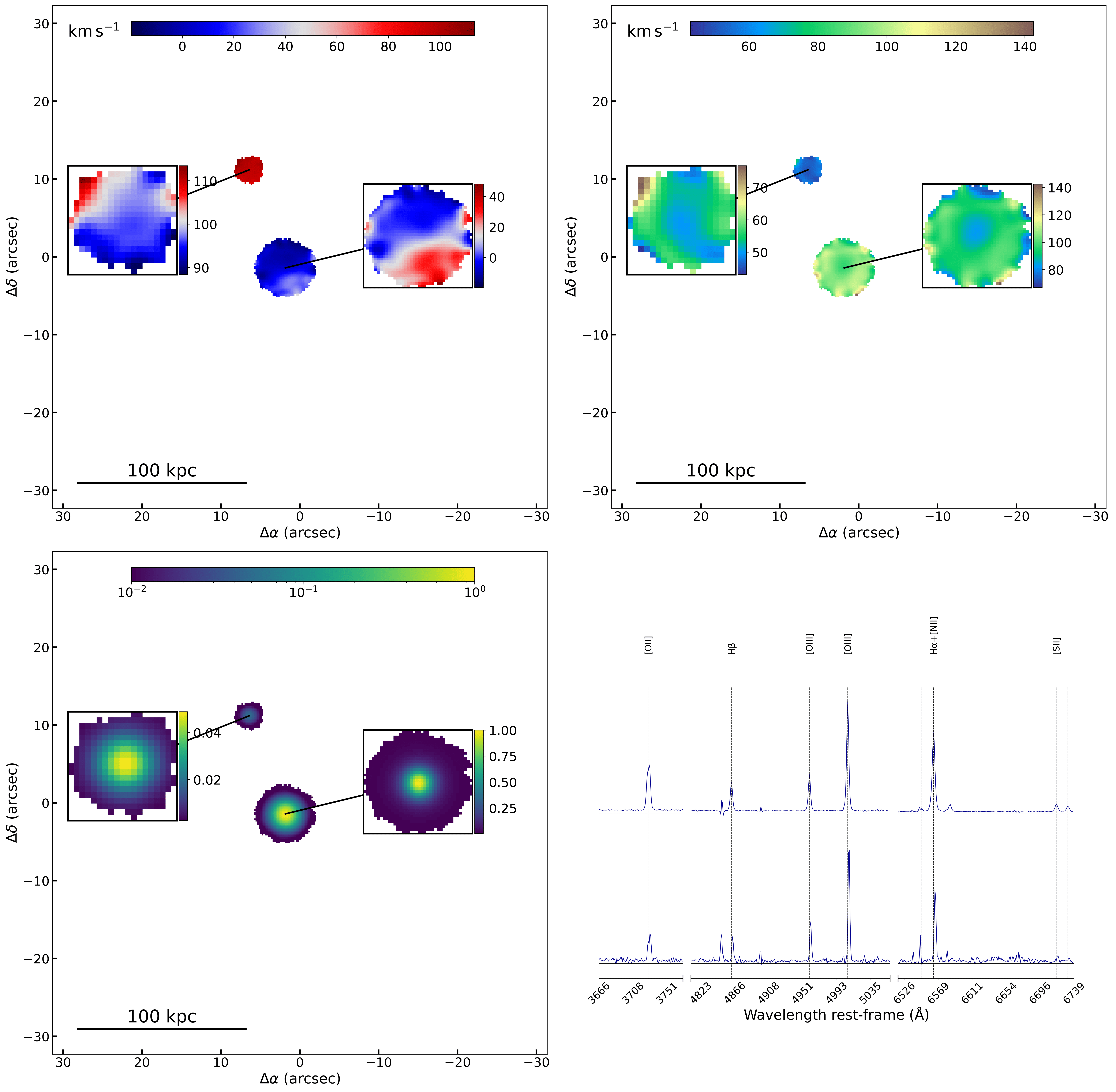}
     \caption{GP23 system. For details of the meaning of each panel, see Fig. \ref{panel_GP1}}
     \label{panel_GP23}
\end{figure*}

\begin{figure*}[h!]
\centering
   \includegraphics[width=\textwidth]{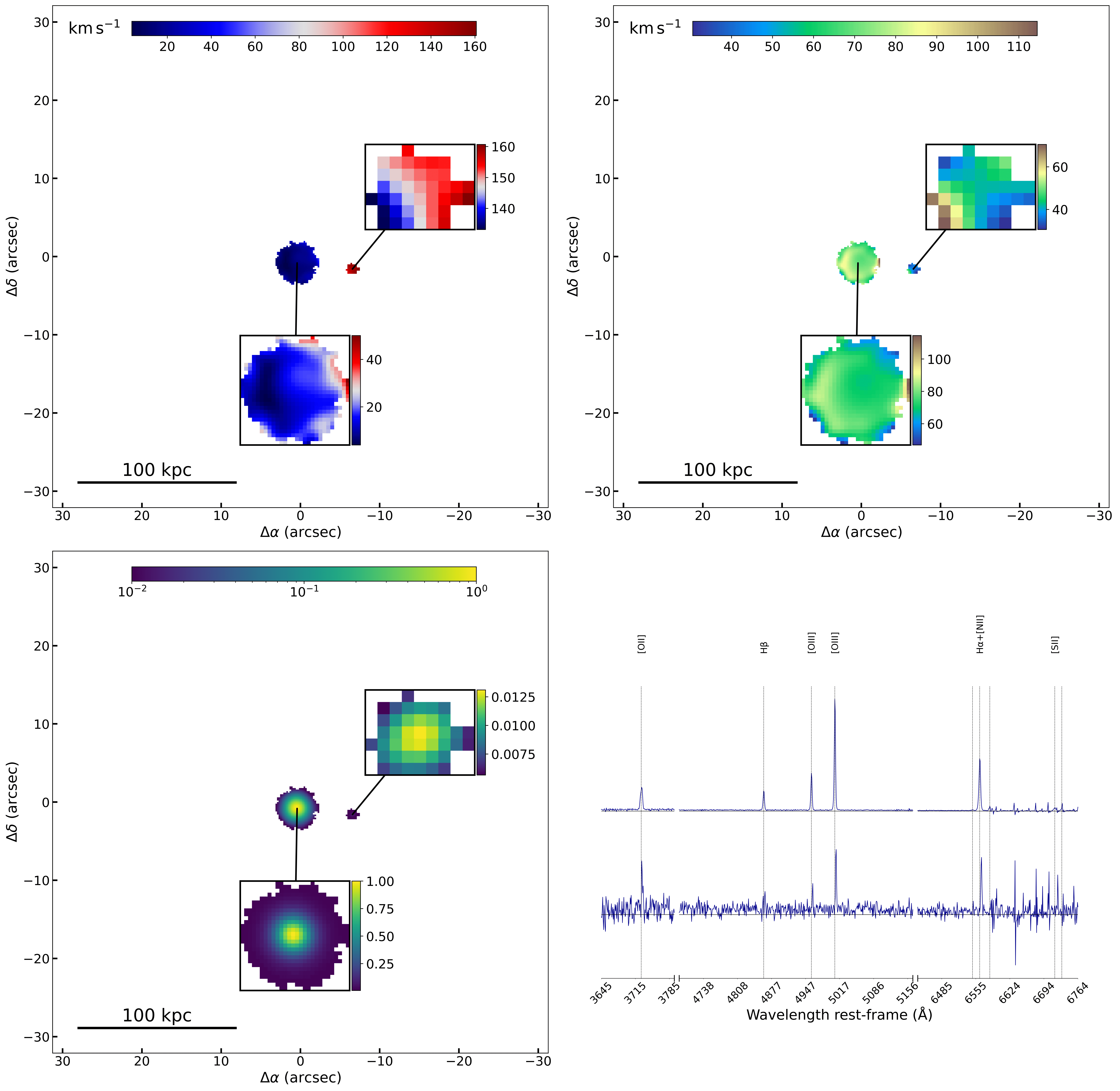}
     \caption{GP24 system. For details of the meaning of each panel, see Fig. \ref{panel_GP1}}
     \label{panel_GP24}
\end{figure*}

\end{appendix}

\end{document}